\documentclass[journal]{IEEEtran}

\usepackage{cite}

\usepackage{graphicx}
\DeclareGraphicsExtensions{.eps}

\usepackage{amssymb}
\usepackage{mathrsfs}
\usepackage{bbm}
\usepackage{cite}
\usepackage{caption}
\usepackage{subcaption}
\usepackage[table]{xcolor}
\usepackage{graphicx}
\usepackage{color}
\usepackage{placeins}
\usepackage{float}
\usepackage{tabularx,colortbl}
\usepackage[cmex10]{amsmath}
\usepackage{amssymb}
\usepackage{amsbsy}
\usepackage{bm}
\usepackage{amsthm}
\usepackage{bbm}
\usepackage{epstopdf}
\usepackage{breqn}
\usepackage{framed}
\usepackage{url}
\usepackage{amsfonts}
\usepackage{mathrsfs}
\usepackage{mathtools}
\usepackage{algorithm}
\usepackage{algpseudocode}
\usepackage{dblfloatfix}
\usepackage{afterpage}

\newcommand*{\QEDB}{\hfill\ensuremath{\blacksquare}}%

\usepackage{url}

\newtheorem{thm}{Theorem}

\newtheorem{prop}[thm]{Proposition}

\begin{document}
%
\title{Dynamic Bayesian Multitaper Spectral Analysis}
%
%
%

\author{Proloy~Das~and~Behtash~Babadi~\IEEEmembership{}
\thanks{The authors are with the Department of Electrical and Computer Engineering, University of Maryland, College Park, MD 20742.}
\thanks{This work has been presented in part at the IEEE Signal Processing in Medicine and Biology Symposium, 2017 \cite{spmb}.}
\thanks{This material is based upon work supported by the National Science Foundation under Grant No. 1552946.}
}




\maketitle

\begin{abstract}
Spectral analysis using overlapping sliding windows is among the most widely used techniques in analyzing non-stationary time series. Although sliding window analysis is convenient to implement, the resulting estimates are sensitive to the window length and overlap size. In addition, it undermines the dynamics of the time series as {the estimate associated to each window} uses only the data within. Finally, the overlap between consecutive windows hinders a precise statistical assessment. {In this paper, we address these shortcomings by explicitly modeling the spectral dynamics through integrating the multitaper method with state-space models in a Bayesian estimation framework. The underlying states pertaining to the eigen-spectral quantities arising in multitaper analysis are estimated using instances of the Expectation-Maximization algorithm, and are used to construct spectrograms and their respective confidence intervals. We propose two spectral estimators that are robust to noise and are able to capture spectral dynamics at high spectrotemporal resolution.} We provide theoretical analysis of the bias-variance trade-off, which establishes performance gains over the standard overlapping multitaper method. We apply our algorithms to synthetic data as well as real data from human EEG and electric network frequency recordings, the results of which validate our theoretical analysis.
\end{abstract}

\begin{IEEEkeywords}
Spectrogram analysis, Non-stationary spectral analysis, Multitaper analysis, {State-space models}, Bayesian filtering
\end{IEEEkeywords}

%
\IEEEpeerreviewmaketitle

\section{Introduction}

\IEEEPARstart{S}{pectral} analysis techniques are among the most important tools for extracting information from time series data recorded {from} naturally occurring processes. Examples include speech \cite{Quatieri:08}, images \cite{lim1990two}, electroencephalography (EEG) \cite{Buzsaki:09}, oceanography \cite{emery2001data}, climatic time series \cite{ghil2002advanced} and seismic data \cite{Yilmaz:01}. Due to the exploratory nature of most of these applications, non-parametric techniques based on Fourier methods and {wavelets} are among the most widely used. In particular, the multitaper (MT) method excels among the available non-parametric techniques due to both its simplicity and {control over the bias-variance trade-off via bandwidth adjustment \cite{mtm,percival1993,bronez}}. 

Most existing spectral analysis techniques assume that the time series is stationary. In many applications of interest, however, the energy of the various oscillatory components in the data exhibits dynamic behavior. {Extensions of stationary time series analysis to these non-stationary processes have led to `time-varying' spectral descriptions such as the Wigner-Ville distribution \cite{Cohen95,martin1985wigner}, the evolutionary spectra and its generalizations \cite{priestley1965evolutionary,matz1997generalized}, and the time-frequency operator symbol formulation \cite{matz2006nonstationary} (See \cite{kozek1996matched} for a detailed review).} A popular approach to {estimating} such time-varying spectra is to subdivide the data into overlapping windows or segments and estimate the spectrum  locally for each window using various Fourier or wavelet-based methods\cite{HAMMOND1996419}, assuming the underlying process is quasi-stationary, i.e., the spectrum changes slowly with time. Thereby, the so-called spectrogram analysis is obtained by using sliding windows with overlap in order to capture non-stationarity.

Although sliding window processing is widely used due to its fast implementation, it has several major drawbacks.  {First, the window length and extent of overlap are subjective choices and can drastically change the overall attribute of the spectrogram if chosen poorly.} Second, given that {the estimate associated to a given window is obtained by only the data within}, it ignores the common dynamic trends shared across multiple windows, and thereby fails to fully capture the degree of smoothness inherent in the signal. Instead, the smoothness of the estimates is enforced by the amount of overlap between adjacent windows. {Third, although techniques such as the MT analysis are able to mitigate the variabilities arising from finite data duration or the so-called `sampling' noise by averaging over multiple tapers, their spectral resolution degrades when applied to data within small windows due to the increase in the Rayleigh resolution \cite{rayleigh1994collected}. In addition, they do not have a mechanism in place to suppress the additive measurement noise that commonly contaminates empirical observations.} {Fourth}, the overlap between adjacent windows hinders a precise statistical assessment of the estimates, such as constructing confidence intervals due to the high dependence of estimates across windows. To address this issue, statistical corrections for multiple comparisons need to be employed \cite{efron1982jackknife}, which in turn limit the resulting test powers when multiple windows are involved.

In recent years, several alternative approaches to non-stationary spectral analysis have been proposed, such as the empirical mode decomposition (EMD) \cite{Huang903full}, synchrosqueezed wavelet transform \cite{daubechies2011synchrosqueezed,daubechies2016conceft}, time-frequency reassignment \cite{xiao2007multitaper}, {time-frequency ARMA models \cite{Jachan2007TFARMA}}, and spectrotemporal pursuit \cite{ba2014}. These techniques aim at decomposing the data into a small number of smooth oscillatory components in order to produce spectral representations that are smooth in time but sparse or structured in frequency. Although they produce spectral estimates that are highly localized in the time-frequency plane, they require certain assumptions on the data to hold. {For example, EMD analysis assumes the signal to be deterministic and does not take into account the effect of observation noise \cite{Huang903full}.} Other methods assume that the underlying spectrotemporal components pertain to certain structures such as amplitude-modulated narrowband mixtures \cite{daubechies2011synchrosqueezed, daubechies2016conceft}, sparsity \cite{ba2014} or chirp-like dynamics \cite{xiao2007multitaper}. In addition, they lack a statistical characterization of the estimates. Finally, although these sophisticated methods provide spectrotemporal resolution improvements, they do not yield implementations as simple as those of the sliding window-based spectral estimators.

{In this paper, we address the above-mentioned shortcomings of {sliding window multitaper estimators} by resorting to state-space modeling. State-space models provide a flexible and natural framework for analyzing systems that evolve with time \cite{fahrmeir2013multivariate, shum,ssm2003,Kitagawa1987}, and have been previously used for parametric \cite{bohlin1977analysis, Jachan2007TFARMA} and non-parametric \cite{ba2014} spectral estimation. The novelty of our approach is in the integration of techniques from MT analysis and state-space modeling in a Bayesian estimation framework. To this end, we construct state-space models in which the underlying states pertain to the eigen-spectral quantities, such as the empirical eigen-coefficients and eigen-spectra arising in MT analysis. We employ state dynamics that capture the  evolution of these quantities, coupled with observation models that {reflect} the effect of {measurement} and sampling noise. We then utilize Expectation-Maximization (EM) to find the maximum \emph{a posteriori} (MAP) estimate of the states given the observed data to construct our spectral estimators as well as statistical confidence intervals.}

{We provide theoretical analysis of the bias-variance trade-off, which reveals two major features of our proposed framework: 1) our methodology inherits the control mechanism of the bias-variance trade-off from the MT framework by means of changing the design bandwidth parameters \cite{bronez}, and 2) our algorithms enjoy the optimal data combining and denoising features of Bayesian filtering and smoothing.} In addition, due to the simplicity and wide usage of Bayesian filtering and smoothing algorithms, our algorithms are nearly as simple to implement as the sliding window-based spectrograms. To further demonstrate the performance of our algorithms, we apply them to synthetic as well as real data including human EEG recordings during sleep and electric network frequency data from audio recordings. Application of our proposed estimators to these data provides spectrotemporal features that are significantly denoised, are smooth in time, and enjoy high spectral resolution, thereby corroborating our theoretical results.

The rest of the paper is organized as follows: In Section \ref{problem}, we present the preliminaries and problem formulation. In Section \ref{sec:algorithms}, we develop our proposed estimators. Application of our estimators to synthetic and real data are given in Section \ref{sec:application}, followed by our theoretical analysis in Section \ref{sec:theory}. Finally, our concluding remarks are presented in Section \ref{sec:conclusion}.

\section{Preliminaries and Problem Formulation}
\label{problem}
\subsection{Non-stationary Processes and Time-Varying Spectrum}
Consider a finite realization of $T$ samples from a discrete-time non-stationary process $y_t, t = 1,2, \cdots T$, obtained via sampling a 
 continuous-time signal above Nyquist rate. We assume that the non-stationary process $y_t$ is harmonizable so that it admits a Cram\'{e}r representation\cite{Loeve63} of the form: 
\begin{equation}
\label{eq:cramer}
y_t = \int_{-\frac{1}{2}}^{\frac{1}{2}} e^{i2\pi f t} dz(f),
\end{equation}
where $dz(f)$ is the generalized Fourier transform of the process. This process has a covariance function of the form:
\begin{align} \label{eq:gen_covar}
\!\!\!\!\resizebox{.94\columnwidth}{!}{$\displaystyle \Gamma_{L}(t_1,t_2)\!:=\!\mathbbm{E}[y_{t_1}y^{*}_{t_2{}}]\!=\!  \int_{-\frac{1}{2}}^{\frac{1}{2}}\!\int_{-\frac{1}{2}}^{\frac{1}{2}} e^{i2\pi(t_1f_1-t_2f_2)}\gamma_{L}(f_1,f_2)df_1df_2$,}  
\end{align}
where $ \gamma_{L}(f_1,f_2)\!\!:=\!\!\mathbbm{E}[dz(f_1)dz^{*}(f_2)] $ is referred to as the generalized spectral density or the Lo\`{e}ve spectrum \cite{thomsonbc}. Due to the difficulty in extracting physically-plausible spectrotemporal information from the two-dimensional function $\gamma_{L}(f_1,f_2)$, other forms of spectrotemporal characterization that are two-dimensional functions over time and frequency have gained popularity \cite{HAMMOND1996419}. To this end, by defining the coordinate rotations $t := (t_1 + t_2)/2$, $\tau := t_1 - t_2$, $f := (f_1 + f_2)/2$, and $g := f_1 -f_2$ and by substituting in the definition of the covariance function in (\ref{eq:gen_covar}), we obtain:  
\begin{align} \nonumber
\resizebox{!}{!}{$\displaystyle \Gamma(\tau,t)\!:=\!\Gamma_{L}(t_1,t_2)\!\!=\!\!\!\int_{-\frac{1}{2}}^{\frac{1}{2}}\!\int_{-\frac{1}{2}}^{\frac{1}{2}}\!\!\!\!e^{i2\pi(tg+\tau f)}\gamma(g,f)dfdg$,} 
\end{align}
where $f$ and $g$ are referred to as the ordinary and non-stationary frequencies, respectively\cite{thomsonbc}, and 
$\gamma(g,f)df dg := \gamma_{L}(f_1,f_2)df_1df_2$ 
is the Lo\`{e}ve spectrum in the rotated coordinates. To obtain one such two-dimensional spectral density representation over time and frequency, we define:
\begin{align} \label{eq:Wigner}
 \resizebox{0.91\columnwidth}{!}{$\displaystyle \!\!\!\!\!D(t,f)\!:=\!\! \int_{-\frac{1}{2}}^{\frac{1}{2}}\!\!e^{i2\pi tg} \gamma(g,f) dg\!=\!\!\int_{-\frac{1}{2}}^{\frac{1}{2}}\!\!e^{-i2\pi\tau f} \mathbbm{E}[y_{t+\frac{\tau}{2}}y_{t-\frac{\tau}{2}}^{*}] d\tau,$}
\end{align}
which coincides with the expected value of the Wigner-Ville distribution \cite{Cohen95}. The `time-varying' spectral representation $D(t,f)$ captures the spectral information of the data as a function of time, and thus provides a useful framework for analyzing non-stationary time series. {However, estimating $D(t,f)$ from finite samples of the process is challenging, considering that the expectation needs to be replaced by time averages which may smooth out the time-varying features of the signal \cite{martin1985wigner}.}

{In order to address this challenge, certain additional assumptions need to be imposed on the underlying process, which as a matter of fact restrict the extent of temporal or spectral variations the signal can exhibit. In this regard, two such popular assumptions are posed in terms of the \emph{quasi-stationary} and \emph{underspread} properties. In order to define these properties quantitatively, let us first define the Expected Ambiguity Function (EAF) \cite{kozek1996matched,matz1997generalized} as:
\begin{align}
\resizebox{0.91\columnwidth}{!}{$\displaystyle \!\!\!\!\!\Delta(\tau,g)\!:=\!\int_{-\frac{1}{2}}^{\frac{1}{2}}\!\!e^{i2\pi \tau f} \gamma(g,f) df\!=\!\! \int_{-\frac{1}{2}}^{\frac{1}{2}}\!\!e^{-i2\pi t g} \mathbbm{E}[y_{t+\frac{\tau}{2}}y_{t-\frac{\tau}{2}}^{*}] dt$}
\end{align}
A signal whose EAF has a limited spread along $g$ (i.e., negligible spectral correlation) is called \emph{quasi-stationary}. If the signal is concentrated around the origin with respect to both $\tau$ and $g$ (i.e., small spectral and temporal correlation), it is called \emph{underspread.} There exists a large body of work on estimating time-varying spectra under the quasi-stationarity assumption, such as short-time periodograms, pseudo-Wigner estimators \cite{martin1985wigner}, and estimates of the evolutionary spectra \cite{priestley1965evolutionary}. More recent methods such as the Generalized Evolutionary Spectra (GES) estimators \cite{matz1997generalized} and time-frequency auto-regressive moving-average (TFARMA) \cite{Jachan2007TFARMA} estimators rely on the underspread property of the underlying signals. We refer the reader to \cite[Chapter 10]{hlawatsch2013time} for a detailed discussion of these properties.}

{It is noteworthy that both assumptions are fairly general, encompass a broad range of naturally-occurring processes, and have resulted in successful applications in real life problems. We will next discuss one of the widely used methods for estimating time-varying spectra under the quasi-stationary assumption, which extends MT spectral analysis beyond second-order stationary processes and has gained popularity in exploratory studies of naturally occurring processes \cite{thomsonbc}.}  

\subsection{The Sliding Window MT Spectral Analysis}
{One of the popular non-parametric techniques for estimating the `time-varying' spectral representation $D(t,f)$ is achieved by subdividing the data into overlapping windows or segments and estimating the spectrum for each window independently using the MT method \cite{thomsonbc}. This method naturally and intuitively extends the popular non-parametric MT method to the non-stationary scenario under the assumption of quasi-stationarity, which enables one to treat the time series within segments (locally) as approximately second-order stationary \cite{priestley1981spectral}. The resulting spectrotemporal representations are smoothed version of the Wigner-Ville distribution \cite{kozek1996matched} and are referred to as  spectrogram. In what follows, we briefly describe the  MT spectrogram method, since it preludes the rest of our treatment.}

The MT method is an extension of single-taper spectral analysis, where the data is element-wise multiplied by a taper prior to forming the spectral representation to mitigate spectral leakage \cite{percival1993, mtm}. In the MT method, spectral representation is computed as the average of several such single-taper PSDs, where the tapers are orthogonal to each other and exhibits good leakage properties. This can be achieved by using the  \textit{discrete prolate spheroidal sequences} (dpss) or {\textit{Slepian sequences}} \cite{slepian78}, due to their orthogonality and optimal leakage properties.

Another viewpoint of the MT method with this particular choice of data tapers is the decomposition of the spectral representation of the process over a set of orthogonal basis functions. Indeed, these basis functions originate from an approximate solution to the integral equation expressing the projection of $dz(f)$ onto the Fourier transform of the data:
\begin{align}
\nonumber y(f) = \int_{-\frac{1}{2}}^{\frac{1}{2}}\frac{\sin W \pi (f-\zeta)}{\sin \pi (f-\zeta)} e^{-i2\pi (f-\zeta) \frac{W-1}{2}}dz(\zeta).
\end{align}
where $W$ is window length, i.e., number of samples, and $dz(\zeta)$ is an orthogonal increment process. This integral equation can be approximated using a local expansion of the increment process over an interval $[-B, B]$, for some small design band-width $B$, in the space spanned by the eigenfunctions of the Dirichlet kernel $\frac{\sin W \pi f}{\sin \pi f}$ \cite{percival1993, mtm}. 

These eigenfunctions are known as the prolate spheroidal wave functions (PSWFs), which are a set of doubly-orthogonal functions over $[-B, B]$ and $[-{1}/{2}, {1}/{2}]$, with time-domain representations given by the dpss sequences. Let $u_l^{(k)}$ be the $l$th sample of the $k$th dpss sequence, for a given bandwidth $B$ and window length $W$. The $k$th PSWF is then defined as:
\vspace{-2mm}
\begin{align}
\nonumber U^{(k)}(f) := \sum_{l = 0}^{W-1} u_l^{(k)}e^{-i2\pi f l}.
\end{align}
{Choosing $K \leqslant \lfloor2WB\rfloor -1$ dpss having eigenvalues close to $1$ as data tapers}, the {MT spectral estimate} can be calculated as follows:   
\begin{align}
\label{eq:mtsa}
\widehat{S}^{(\sf mt)}(f) := \frac{1}{K}\sum_{k=1}^{K}|{\widehat{x}^{(k)}(f)}|^2, 
\end{align}
where ${\widehat{x}^{(k)}(f)} := \sum_{l=0}^{W-1}e^{-i2\pi f l}u_l^{(k)}y_{l}$ for $k=1,2,\cdots, K$ are called the `eigen-coefficients'. The `eigen-spectra', $\widehat{S}^{(k)}(f) := |{\widehat{x}^{(k)}(f)}|^2$ can be viewed as the expansion coefficients of the decomposition. 
 
To estimate time-varying spectra under the MT framework, sliding windows with overlap are used to enforce temporal smoothness and increase robustness of the estimates \cite{reviewmtm,thomsonbc}, resulting in MT spectrogram estimates. {Although this `overlapping' MT procedure overcomes frequency leakage issues and produces consistent estimates, subjective choices of the window length and the degree of overlap can change the overall appearance of the spectrogram drastically when poor choices of these parameters are used. }
In addition, these estimates lack precise statistical inference procedures, such as hypothesis testing, due to the statistical dependence induced by the overlaps. The objective of this work is to overcome these limitations by directly modeling and estimating the evolution of the process without committing to overlapping sliding windows, while achieving fast and efficient implementations. {Before presenting our proposed solutions, we give a brief overview of other existing approaches in the literature in order to put our contributions in context.}

\vspace{-2mm}
\subsection{Motivation and Connection to Existing Literature}
{Our goal is to overcome the foregoing challenges faced by the sliding window MT spectrogram analysis in estimating the expected value of the Wigner-Ville distribution (See Eq. (\ref{eq:Wigner})) under the quasi-stationarity assumption. In addition, our work can be viewed in the context of the spectrogram approximation to the evolutionary spectra \cite{priestley1965evolutionary}. The evolutionary spectra is obtained by considering an expansion of $y_t$ over the set of complex sinusoids as 
\begin{equation}
\displaystyle y_t = \int_{-1/2} ^{1/2}x_t(f)e^{j2\pi ft} df,
\end{equation}
 with uncorrelated time-varying expansion coefficients, i.e., $\displaystyle \mathbbm{E}[x_t(f_1)x_t^{*}(f_2)] = D(t,f_1)\delta(f_1-f_2)$. In standard MT spectrogram analysis, the extent of overlap between consecutive segments dictates the amount of temporal smoothness in the estimates. Our approach is to avoid the usage of overlapping windows by modeling and estimating the dependence of the spectra across windows using state-space models, while retaining the favorable leakage properties of the MT analysis. As will be revealed in the subsequent sections, in the same vein as the sliding window multitaper analysis our methods pertain to the class of spectrogram estimates, which are viewed as smoothed versions of the Wigner-Ville spectrum \cite{kozek1996matched}.} 

{Due to the underlying quasi-stationarity assumption, i.e., negligible spectral correlation, the domain of applicability of our methods might be narrower than the more general non-stationary spectral analysis methods such as GES and Weyl spectral estimation and TFARMA modeling; however, our methods admit simple and efficient implementations, which makes them attractive for exploratory applications in which sliding window processing is widely used with subjective and ad hoc choices of design parameters. In this context, the novelty of our contributions lies in:}
\begin{itemize}
\item[1)] Capturing the evolution of the spectra across windows by modeling the dynamics of certain eigen-spectral quantities arising in MT analysis (e.g., spectral eigen-coefficients and eigen-spectra);   

\item[2)] Addressing the additive measurement noise and multiplicative sampling noise, which severely distort the spectrograms obtained by the multitaper framework; and

\item[3)] Constructing a framework for precise statistical assessment of the estimates, by addressing the dependency among windows using a Bayesian formulation.   
\end{itemize} 

As it will be evident in Section \ref{sec:theory}, the use of state-space models in the context of MT analysis results in adaptive weighting of the estimates of the eigen-coefficients or eigen-spectra across windows, thanks to the optimal data combining feature of Bayesian smoothing. These adaptive weights depend on the common dynamic trends shared across windows and hence result in capturing the degree of smoothness inherent in the signal, while producing estimates robust against uncertainties due to observation noise and limited data.

{It is noteworthy that the use of state-space models in our work is significantly different from those  used in \emph{parametric} non-stationary spectral analysis methods such as TFARMA modeling \cite{Jachan2007TFARMA}. In the TFARMA formulation, time delays and frequency shifts are used to model the non-stationary dynamics of the process in a physically intuitive way. These state-space models therefore determine the functional form of the resulting spectral estimates in closed form in terms of the finite set of ARMA coefficients. The state-space models used in our work, however, do not determine the functional form of the spectral estimates at each window, and rather control the temporal smoothness of the eigen-spectral quantities via forming a regularization mechanism in the underlying Bayesian estimation framework (See Section \ref{sec:inverse}). In our approach, we indeed estimate the spectrogram at a given number of frequency bins in each window, which scales with the total number of samples, in the same vein as sliding window MT spectrogram.}

{In light of the above, our algorithms belong to the class of \emph{semi-parametric} estimation methods, as the underlying model is a hybrid of \emph{parametric} unobservable state evolution process and a \emph{non-parametric} data generating process \cite{powell1994estimation}. In Section \ref{sec:simulation}, we will compare our proposed semi-parametric methodology with both non-parametric and parametric techniques, namely the MT spectrogram analysis and the Time-Frequency Autoregressive (TFAR) modeling technique \cite{Jachan2007TFARMA}.}
\color{black}
\vspace{-2mm}  
\subsection{Problem Formulation}

Assume, without loss of generality, that an arbitrary window of length $W$ is chosen so that for some integer $N$, $NW = T$ and let $\mathbf{y}_n = \big[y_{(n-1)W+1}, y_{(n-1)W+2} ,\cdots, y_{nW}\big]^\top$ for $n = 1,2 \cdots N$, denotes the data in the $n$th window. This way, the entire data is divided into $N$ non-overlapping segments of length $W$ each. To this end, we invoke quasi-stationarity assumption by modeling $y_t$ to be stationary within each of these segments of length $W$. With this assumption, motivated by the major sources of uncertainty in spectral estimation, i.e., measurement noise and sampling noise, we formulate two state-space frameworks in the following subsections.  
 
\subsubsection{Mitigating the Measurement Noise} 
\label{sec:MeasurementNoise}
 {Suppose that $\widetilde{y}_t$ is the noise corrupted observation obtained from the true signal $y_t$, i.e., $\widetilde{y}_t = y_t + v_t$, where $(v_t)_{t=1}^T$ is an i.i.d. zero-mean Gaussian noise sequence with fixed variance $\sigma^2$}. By discretizing the representation in (\ref{eq:cramer}) at a frequency spacing of $2\pi/J$ with $J$ an integer, at any arbitrary window $n$, we have
 \begin{align}
 \label{eq:meas_noise_model}
 \widetilde{\mathbf{y}}_n = \mathbf{F}_n \mathbf{x}_n + \mathbf{v}_n,
 \end{align} 
 where $\mathbf{F}_n$ is a matrix with elements $(\mathbf{F}_n)_{l,j} := \exp\big(i2\pi(((n-1)W+l)\frac{j-1}{J})\big)$ for $l = 1,2,\cdots, W$ and $j = 1,2,, \cdots J$; $\widetilde{\mathbf{y}}_n := \big[\widetilde{y}_{(n-1)W+1}, \widetilde{y}_{(n-1)W+2} ,\cdots, \widetilde{y}_{nW}\big]^\top$ is the noisy observation of the true signal $\mathbf{y}_n$; $x_n(f)$ and $\mathbf{x}_n:= \big[x_n(0), x_n(2\pi\frac{1}{J}), \cdots ,x_n(2\pi\frac{J-1}{J})\big]^\top$ denote the orthogonal increment process and its discretized version, respectively at window $n$ and $\mathbf{v}_n = \big[v_{(n-1)W+1}, v_{(n-1)W+2} ,\cdots, v_{nW}\big]^\top$ is zero-mean Gaussian noise with covariance $\text{Cov}\{\mathbf{v}_i,\mathbf{v}_j\} = \sigma^2\mathbf{I}\delta_{i,j}$ .
 
Let $ \mathbf{u}^{(k)}\!:=\!\big[u_{1}^{(k)}, u_{2}^{(k)} ,\cdots, u_{W}^{(k)}\big]^\top $ denotes the $k$th dpss taper and {$\widetilde{\mathbf{y}}_n^{(k)} := \mathbf{u}^{(k)} \odot \widetilde{\mathbf{y}}_n$}, where $\odot$ denotes element-wise multiplication.  Let $x_n^{(k)}(f)$ and $\mathbf{x}_n^{(k)} := \left [x_n^{(k)}(0), x_n^{(k)}(2\pi\frac{1}{J}), \cdots ,x_n^{(k)}(2\pi\frac{J-1}{J})\right ]^\top$ denote the $k$th spectral eigen-coefficient of $\mathbf{y}_n$ and its discretized version, respectively, for $k=1,2,\cdots,K$. Then, following (\ref{eq:meas_noise_model}) we consider the following spectrotemporal representation of the tapered data segments: 
\vspace{-2mm}
\begin{align}
\label{eq:formod1}
	 \widetilde{\mathbf{y}}_n^{(k)}= \mathbf{F}_n \mathbf{x}_{n}^{(k)} +  \mathbf{v}_n^{(k)},
\end{align}
where $\mathbf{v}^{(k)}_n$ is the contribution of $\mathbf{v}_n$ to the $k$th tapered data, assumed to be independent of $\mathbf{x}_{1:n-1}^{(k)}$, and identically distributed according to a zero-mean Gaussian distribution with covariance $\text{Cov}\{\mathbf{v}^{(k)}_i,\mathbf{v}^{(k)}_j\} ={\sigma^{(k)}}^2 \mathbf{I}\delta_{i,j}$. We view $\widetilde{\mathbf{y}}_n^{(k)}$ as a noisy observation corresponding to the true eigen-coefficient $\mathbf{x}_{n}^{(k)}$, which provides a linear Gaussian forward model for the observation process.   

In order to capture the evolution of the spectrum and hence systematically enforce temporal smoothness, we impose a stochastic continuity constraint on the eigen-coefficients $( \mathbf{x}_{n}^{(k)})_{n=1}^{N}$ for $ k = 1,2, \cdots K$, using a first-order difference equation:
\vspace{-2mm}
\begin{align} \label{eq:scc1} 
    \mathbf{x}_n^{(k)} = \alpha^{(k)} \mathbf{x}_{n-1}^{(k)} + \mathbf{w}^{(k)}_n,
\end{align}%
where $0 \leqslant \alpha^{(k)}<1$, and $\mathbf{w}^{(k)}_n$ is independent of $\mathbf{x}_{1:n-1}^{(k)}$ and assumed to be independently distributed according to a zero-mean Gaussian distribution with diagonal covariance $\text{Cov}\{\mathbf{w}^{(k)}_i,\mathbf{w}^{(k)}_j\} = \mathbf{Q}^{(k)}_i \delta_{i,j}$. Under this assumption, the discrete-time process, $(\mathbf{x}_n^{(k)})_{n=1}^{N}$ forms a jointly Gaussian random process with independent increments, while the process itself is statistically dependent. An estimate of the unobserved states (true eigen-coefficients) from the observations (tapered data) under this model suppresses the measurement noise and captures the state dynamics.

\subsubsection{Mitigating the Sampling Noise}
\label{sec:SamplingNoise}
Suppose the additive measurement noise is negligible, i.e. $v_t \approxeq 0$. For now, consider only a single window of length $W$. It is known that when the spectrum does not rapidly vary over the chosen design bandwidth $B$, the eigen-spectra are approximately uncorrelated and the following approximation holds for $k = 1,2 \cdots, K$ \cite{thomsonbc,percival1993}:
\begin{align}\label{eq:chi}
\frac{\widehat{S}^{(k)}(f)}{S(f)} \sim \frac{\chi_{2}^2}{2}, \text{    } 0 <f <1/2,
\end{align}
where $\widehat{S}^{(k)}(f)$ and $S(f)$ are the tapered estimate and true PSD, respectively. {In other words, the empirical eigen-spectra of the process can be thought of as the true spectra corrupted by a multiplicative noise, due to sampling and having access to only a single realization of the process. We refer to this uncertainty induced by sampling as \emph{sampling noise}.} {By defining  $\psi^{(k)}(f) := \log{\widehat{S}^{(k)}(f)} + \log2$ and $s^{(k)}(f) := \log{S(f)}$, we can transform the multiplicative effect of the {sampling noise} in Eq. (\ref{eq:chi}) to the following additive forward model\cite{rosen2009}:
\begin{align} 
\psi^{(k)}(f) = s^{(k)}(f) + \phi^{(k)}(f),
\end{align}    
where $\phi^{(k)}(f)$ is a log-chi-square distributed random variable, capturing the uncertainty due to sampling noise.} It can be shown that $\phi^{(k)}(f)$ has a density given by:
\begin{align}
p(\phi) = \frac{1}{2}\exp\left (\phi-\frac{1}{2}\exp(\phi)\right),
\end{align} 
which belongs to the family of log-Gamma distributions, including the Gumbel and Bramwell-Holdsworth-Pinton distributions common in extreme value statistics \cite{Prentice74}.

In order to incorporate this observation model in our dynamic framework, we define the state vector,  $\mathbf{s}_n^{(k)}\!:=\!\Big[ s_n^{(k)}(0), s_n^{(k)}\Big(\textstyle 2\pi\frac{1}{J}\Big), \cdots ,s_n^{(k)}\Big(\textstyle2\pi\frac{J-1}{J}\Big)\Big]^{\top}$; the observation vector, $\boldsymbol{\psi}_n^{(k)}\!:=\!\Big [\psi_n^{(k)}(0), \psi_n^{(k)}\Big(\textstyle2\pi\frac{1}{J}\Big), \cdots ,\psi_n^{(k)}\Big(\textstyle2\pi\frac{J-1}{J}\Big)\Big]^{\top}$ and the observation noise vector, $
\boldsymbol{\phi}_n^{(k)} := \Big[\phi_n^{(k)}(0), \phi_n^{(k)}\Big(\textstyle2\pi\frac{1}{J}\Big), \cdots ,\phi_n^{(k)}\Big(\textstyle2\pi\frac{J-1}{J}\Big)\Big]^{\top}$. Then, the forward model at window $n$ can be stated as:
\begin{align}\label{eq:formod2}
\boldsymbol{\psi}_n^{(k)} = \mathbf{s}_n^{(k)} + \boldsymbol{\phi}_n^{(k)},
\end{align}
where each element of $\boldsymbol{\phi}_n^{(k)}$ is log-chi-square distributed. Similar to the preceding model, we impose a stochastic continuity constraint over the logarithm of the eigen-spectra as follows:  
\begin{align} \label{eq:scc2}
\mathbf{s}_n^{(k)} = \theta^{(k)} \mathbf{s}_{n-1}^{(k)} + \mathbf{e}_n^{(k)},
\end{align} 
where $0 \leqslant \theta^{(k)} < 1$, and $\mathbf{e}_n^{(k)}$ is assumed to be a zero-mean Gaussian vector {independent of $\mathbf{s}_{1:n-1}^{(k)}$ and} with a diagonal covariance $\text{Cov}\{\mathbf{e}^{(k)}_i,\mathbf{e}^{(k)}_j\} = \mathbf{R}^{(k)}_i \delta_{i,j}$. Note that the logarithm function maps the range of the eigen-spectra in $[0,\infty)$ to $(-\infty,\infty)$ which makes the Gaussian state evolution plausible. An estimate of the unobserved states (logarithm of the true spectra) from the observations (logarithm of the empirical eigen-spectra) under this model suppresses the sampling noise and captures the state dynamics.

\begin{figure}[!t]
\centering
\includegraphics[width=.75\columnwidth]{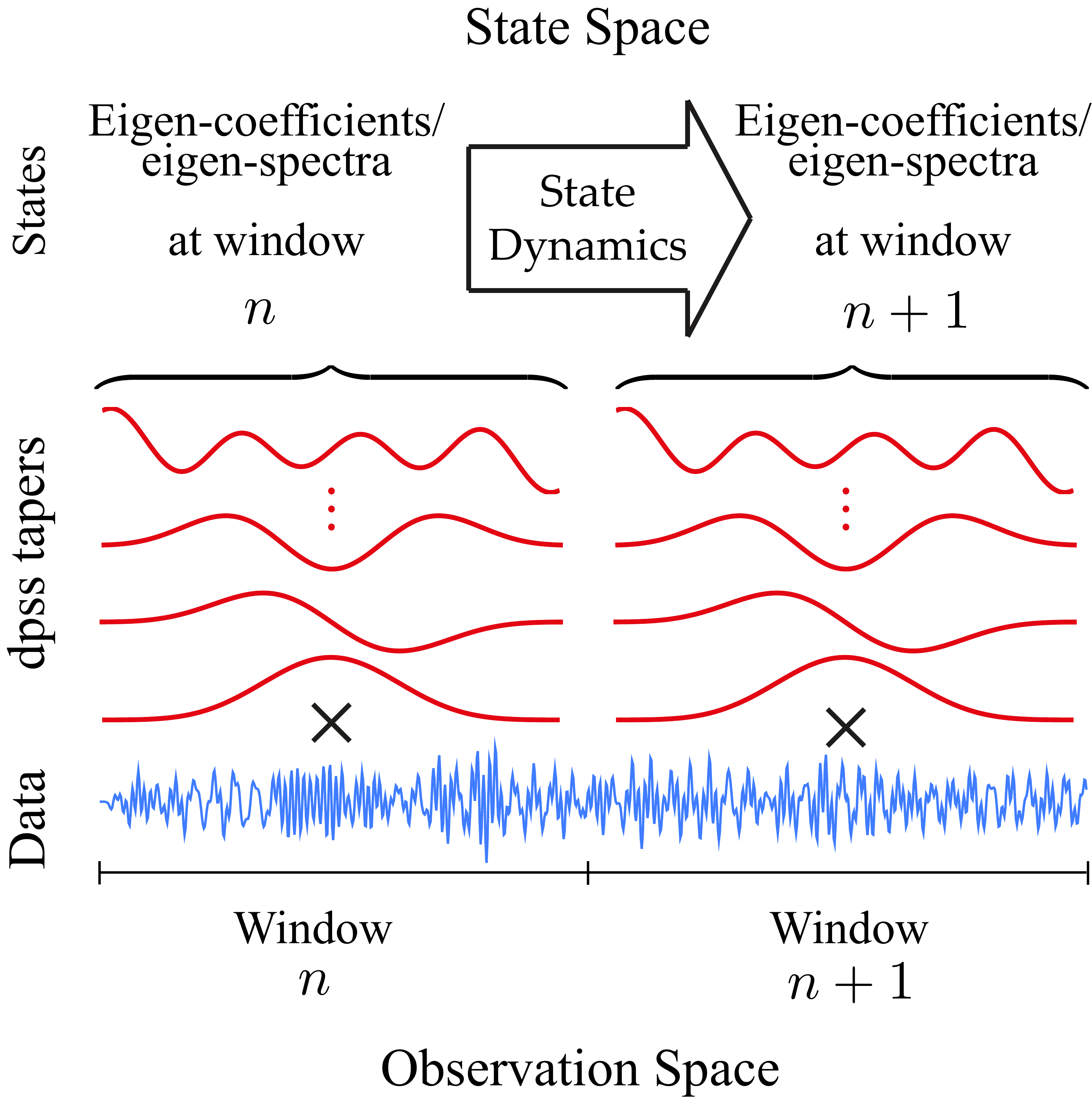}%
\caption{Schematic depiction of the proposed models.}
\label{fig:scheme}
\vspace{-5mm}
\end{figure}

In summary, through these models we project the data of each short window onto the functional space spanned by the PSWFs and impose stochastic continuity constraints ((\ref{eq:scc1}) and (\ref{eq:scc2})) on these projections (eigen-coefficients or eigen-spectra) in order to recover spectral representations that are smooth in time and robust against measurement or sampling noise.  Fig. \ref{fig:scheme} provides a visual illustration of the proposed modeling paradigm.  

\vspace{-2mm}
\subsection{The Inverse Problem}\label{sec:inverse}
We formulate the spectral estimation problem as one of Bayesian estimation, in which the Bayesian risk/loss function, fully determined by the posterior density of $(\mathbf{x}_n^{(k)})_{n=1,k=1}^{N,K}$ (resp. $(\mathbf{s}_n^{(k)})_{n=1,k=1}^{N,K}$) given the observations $(\widetilde{\mathbf{y}}_n^{(k)})_{n=1,k=1}^{N,K}$ (resp. $(\boldsymbol{\psi}_n^{(k)})_{n=1,k=1}^{N,K}$) is minimized. We first consider the forward model of (\ref{eq:formod1}), which provides the observed data likelihood given the states. Under the state-space model of Eq. (\ref{eq:scc1}), the $k$th eigen-coefficient can be estimated by solving the following maximum \emph{a posteriori} (MAP) problem:
\begin{align}
\label{opt:DBMT}
\hspace*{-2mm}\min_{\mathbf{x}_1^{(k)},\mathbf{x}_2^{(k)},\cdots,\mathbf{x}_N^{(k)}} &\sum_{n = 1}^{N}\bigg[ \frac{1}{\sigma^2} \left \| \widetilde{\mathbf{y}}_n^{(k)} -\mathbf{F}_n\mathbf{x}^{(k)}_n\right \|_2^2 + \nonumber \\  &\!\!\!\!(\mathbf{x}_n^{(k)} -\alpha\mathbf{x}_{n-1}^{(k)})^H {\mathbf{Q}^{(k)}_{n}}^{-1}(\mathbf{x}_n^{(k)} -\alpha\mathbf{x}_{n-1}^{(k)})\bigg],
\end{align}
for $k = 1,2, \cdots, K$. {Similarly}, in the second state space framework (\ref{eq:formod2}), the eigen-spectra can be obtained by solving another MAP estimation problem:
\begin{align}
\label{opt:log-DBMT}
    \min_{\mathbf{s}_1^{(k)},\mathbf{s}_2^{(k)},\cdots,\mathbf{s}_N^{(k)}} &\sum_{n = 1}^{N}\bigg[ \mathbf{1}_J^\top[\mathbf{s}_n^{(k)} - \boldsymbol{\psi}_n^{(k)} + \frac{1}{2}\exp(\boldsymbol{\psi}_n^{(k)}-\mathbf{s}_n^{(k)})] \nonumber \\  
    +&(\mathbf{s}_n^{(k)} -\theta \mathbf{s}_{n-1}^{(k)})^H {\mathbf{R}_{n}^{(k)}}^{-1}(\mathbf{s}_n^{(k)} -\theta\mathbf{s}_{n-1}^{(k)})\bigg],
\end{align}
for $k = 1,2, \cdots, K$, where $\mathbf{1}_J$ is the vector of all ones of length $J$. We call the MAP estimation problems in (\ref{opt:DBMT}) and (\ref{opt:log-DBMT}) the Dynamic Bayesian Multitaper ({\sf DBMT}) and the {\sf log-DBMT} estimation problems, respectively. Similarly, the respective spectrogram estimates will be denoted by the {\sf DBMT} and {\sf log-DBMT} estimates. 
   
Equation (\ref{opt:DBMT}) is a strictly convex function of $\mathbf{x}^{(k)}_n \in \mathbb{C}^{W}$ and $\mathbf{Q}^{(k)}_n \in \mathbb{S}^W_{++}$ for $n = 1,2\cdots,N$, which can be solved using standard optimization techniques. However, these techniques do not scale well with the data length $N$. A careful examination of the log-posterior reveals a block tri-diagonal structure of the Hessian, which can be used to develop efficient recursive solutions that exploit the temporal structure of the problem. A similar argument holds for the optimization problem in (\ref{opt:log-DBMT}). However, the parameters of these state-space models need to be estimated from the data. In the next section, we show how the EM algorithm can be used to both estimate the parameters and states efficiently from the optimization problems (\ref{opt:DBMT}) and (\ref{opt:log-DBMT}).

\vspace{-3mm}      
\section{Fast Recursive Solutions via the EM Algorithm}\label{sec:algorithms}
In order to solve the MAP problem in (\ref{opt:DBMT}), we need to find the parameters $\mathbf{Q}_{n}^{(k)} \in \mathbb{S}^W_{++}$ and $\alpha^{(k)} \in (0,1]$ for $n = 1,2\cdots,N$ and $k=1,2,\cdots,K$. Similarly $\mathbf{R}_{n}^{(k)} \in \mathbb{S}^W_{++}$ and $\theta^{(k)} \in (0,1]$ need to be estimated for the problem in (\ref{opt:log-DBMT}). If the underlying states were known, one could further maximize the log-posterior with respect to the parameters. This observation can be formalized in the EM framework \cite{em1977,ssm2003,shum}. To avoid notational complexity, we drop the dependence of the various variables on the taper index $k$ in the rest of this subsection. 

\subsection{The {\sf DBMT} Spectrum Estimation Algorithm}
By treating $\mathbf{x}_n, n = 1,2,\cdots,N$ as the hidden variables and $\alpha, \mathbf{Q}_{n},n = 1,2,\cdots,N$ as the unknown parameters to be estimated, we can write the complete log-likelihood as:
\vspace{-2mm}
\begin{align}\label{eq:complete-ll1}
\log L(\alpha,\mathbf{Q}_{1:N}) &:= - \sum_{n = 1}^{N}\!\bigg[\frac{1}{\sigma^2}\left \|\widetilde{\mathbf{y}}_n - \mathbf{F}_n\mathbf{x}_n \right \|_2^2  + \log |\det \mathbf{Q}_n|\nonumber \\[-2mm]
& \!\!\!\!\!\!\!\!\!\!\!\!\!+ (\mathbf{x}_n-\alpha\mathbf{x}_{n-1})^H\mathbf{Q}_n^{-1}(\mathbf{x}_n - \alpha\mathbf{x}_{n-1}) \bigg ] + {c},
\end{align}
\noindent {where $c$ represents the terms that do not depend on $\alpha$, $( \mathbf{Q}_{n})_{n=1}^N$ or $(\mathbf{x}_{n})_{n=1}^N$}. For simplicity of exposition, we assume that $\mathbf{Q}_{n} = \mathbf{Q}$ for $n = 1,2,\cdots,N$. The forthcoming treatment can be extended to the general case with little modification. Also, note that $\sigma^2$ can be absorbed in $\mathbf{Q}$, and thus is assumed to be known. At the $l$th iteration, we have:

\subsubsection{E-Step} 
Given $\alpha^{[l]}, \mathbf{Q}^{[l]}$, for $ n = 1,2,\cdots,N$, the expectations, 
		$\mathbf{x}_{n|N}:=\mathbbm{E}[\mathbf{x}_n|\widetilde{\mathbf{y}}_{1:N},\alpha^{[l]},\mathbf{Q}^{[l]}],$
	$\boldsymbol{\Sigma}_{n|N}:=\mathbbm{E}[(\mathbf{x}_n-\mathbf{x}_{n|N})(\mathbf{x}_n-\mathbf{x}_{n|N})^H|\widetilde{\mathbf{y}}_{1:N},\alpha^{[l]},\mathbf{Q}^{[l]}],$
	$\boldsymbol{\Sigma}_{n,n-1|N}:=\mathbbm{E}[(\mathbf{x}_n\!-\!\mathbf{x}_{n|N})(\mathbf{x}_{n-1}\!-\!\mathbf{x}_{n-1|N})^H|\widetilde{\mathbf{y}}_{1:N},\alpha^{[l]},\mathbf{Q}^{[l]}],$
	can be calculated using the Fixed Interval Smoother (\textit{FIS}) \cite{rauch65} (lines 4 and 5) and the state-space covariance smoothing algorithm \cite{Jong1988} (line 6). These expectations can be used to compute the expectation of the complete data log-likelihood $\mathbbm{E}\big[\log L(\alpha,\mathbf{Q})|\widetilde{\mathbf{y}}_{1:N}, \alpha^{[l]},\mathbf{Q}^{[l]}\big]$.

\begin{algorithm}[!t]
\caption{The {\sf DBMT} Estimate of the $k$th Eigen-coefficient}
\label{alg:DS3}
\begin{algorithmic}[1]
\State Initialize: observations $\widetilde{\mathbf{y}}_{1:N}^{(k)}$; initial guess ${\mathbf{x}^{(0)}_{1:N}}$; initial guess $\mathbf{Q}^{[0]}$; initial conditions $\boldsymbol{\Sigma}_{0|0}$; tolerance ${\sf tol} \in (0,10^{-3})$, Maximum Number of iteration $L_{\max} \in \mathbbm{N}^+$.
\Repeat
\State $l =0$ .
\State Forward filter for $n=1,2,\cdots, N$:
\vspace{-2.5mm}
\begin{small}
\begin{align*}
    &\mathbf{x}_{n|n-1} = \alpha^{[l]} \mathbf{x}_{n-1|n-1}\\
    &{\mathbf{\Sigma}_{n|n-1} = {\alpha^{[l]}}^2\mathbf{\Sigma}_{n-1|n-1} + \mathbf{Q}^{[l]}}\\
   	&\mathbf{K}_n = \mathbf{\Sigma}_{n|n-1}\mathbf{F}_n^H(\mathbf{F}_n\mathbf{\Sigma}_{n|n-1}\mathbf{F}_n^H + \sigma^2\mathbf{I})^{-1}\\
    &\mathbf{x}_{n|n} = \mathbf{x}_{n|n-1} + \mathbf{K}_n(\widetilde{\mathbf{y}}_n - \mathbf{F}_n\mathbf{x}_{n|n-1})\\
    &\mathbf{\Sigma}_{n|n} = \mathbf{\Sigma}_{n|n-1} - \mathbf{K}_n\mathbf{F}_n\mathbf{\Sigma}_{n|n-1}
\end{align*} 
\end{small}
\vspace{-6.5mm}
\State \resizebox{.78\columnwidth}{!}{Backward smoother for $n=N-1,N-2,\cdots,1$}:
\vspace{-2.5mm}
\begin{small}
\begin{align*}
&\mathbf{B}_n = \alpha^{[l]}\mathbf{\Sigma}_{n|n}\mathbf{\Sigma}_{n+1|n}^{-1}\\
&\mathbf{x}_{n|N} = \mathbf{x}_{n|n} + \mathbf{B}_n(\mathbf{x}_{n+1|N} - \mathbf{x}_{n+1|n})\\
&\mathbf{\Sigma}_{n|N} = \mathbf{\Sigma}_{n|n} + \mathbf{B}_n(\mathbf{\Sigma}_{n+1|N}-\mathbf{\Sigma}_{n+1|n})\mathbf{B}_n^H
\end{align*}
\end{small}
\vspace{-5.5mm}
\State \resizebox{.82\columnwidth}{!}{Covariance smoothing for $n=N-1,N-2,\cdots,1$}:
\vspace{-2.5mm}
\begin{small}
\begin{align*}
\hspace{-25.5mm}\mathbf{\Sigma}_{n,n-1|N} = \mathbf{B}_{n-1}\mathbf{\Sigma}_{n|N}
\end{align*}
\end{small}
\State Let {\small $\widehat{\mathbf{X}}^{[l]} := [\mathbf{x}_{1|N}^{H},\mathbf{x}_{2|N}^{H},\cdots,\mathbf{x}_{N|N}^{H}]^{H}$}.
\State Update $\alpha^{[l+1]}$ and $\mathbf{Q}^{[l+1]}$ as:
\vspace{-1mm}
\begin{align*}
\resizebox{.9\columnwidth}{!}{$\displaystyle \alpha^{[l+1]} = \frac{\sum_{n = 2}^{N} \text{Tr} (\boldsymbol{\Sigma}_{n,n-1|N}{\mathbf{Q}^{[l]}}^{-1})+\mathbf{x}_{n-1|N}^H \mathbf{\mathbf{Q}^{[l]}}^{-1}\mathbf{x}_{n|N}}{\sum_{n = 2}^{N} \text{Tr}(\boldsymbol{\Sigma}_{n-1|N}{\mathbf{Q}^{[l]}}^{-1})+\mathbf{x}_{n-1|N}^H {\mathbf{Q}^{[l]}}^{-1}\mathbf{x}_{n-1|N}}$},
\end{align*}
\vspace{-5mm}
\begin{small}
\begin{align*}
&\resizebox{.9\columnwidth}{!}{$\displaystyle \mathbf{Q}^{[l+1]}\!=\frac{1}{N}\sum_{n=1}^{N} \big[\mathbf{x}_{n|N}\mathbf{x}_{n|N}^{H}\!+\!\boldsymbol{\Sigma}_{n|N}\!+\!{\alpha^{[l+1]}}^2 (\mathbf{x}_{n-1|N}\mathbf{x}_{n-1|N}^{H}\!+\!\boldsymbol{\Sigma}_{n-1|N})$} \nonumber \\[-4pt]
 &\qquad \qquad \qquad \resizebox{.7\columnwidth}{!}{$\displaystyle -\alpha^{[l+1]} (\mathbf{x}_{n-1|N}\mathbf{x}_{n|N}^{H}\!+\!\mathbf{x}_{n|N}\mathbf{x}_{n-1|N}^{H}\!+\!2\boldsymbol{\Sigma}_{n,n-1|N}) \big]$}.   
\end{align*}
\end{small}
\vspace{-6mm}
\State Set $l \leftarrow l+1$.
\vspace{1mm}
\Until{$\frac{\|\widehat{\mathbf{X}}^{[l]}-\widehat{\mathbf{X}}^{(l-1)}\|_2}{\|\widehat{\mathbf{X}}^{[l]}\|_2} < {\sf tol}$ or $l = L_{\max}$.}
\State Output: Denoised eigen-coefficients {$\widehat{\mathbf{X}}^{[L]}$} where $L$ is the index of the last iteration of the algorithm, and error covariance matrices $\boldsymbol{\Sigma}_{n|N}$ for $n = 1,2,\cdots,N$ in from last iteration of the algorithm.
\end{algorithmic}
\end{algorithm}
\subsubsection{M-Step}
The parameters for subsequent iterations, $\alpha^{[l+1]}$ and $\mathbf{Q}^{[l+1]}$ can be obtained by maximizing the expectation of (\ref{eq:complete-ll1}). Although this expectation is convex in $\alpha$ and $\mathbf{Q}$ individually, it is not a convex function of both. Hence, we perform cyclic iterative updates for $\alpha^{[l+1]}$ and $\mathbf{Q}^{[l+1]}$ given by:
\vspace{-1mm}
\begin{align}
\resizebox{.80\columnwidth}{!}{$\displaystyle \alpha^{[l+1]} = \frac{\sum_{n = 2}^{N} \text{Tr} (\boldsymbol{\Sigma}_{n,n-1|N}{\mathbf{Q}^{[l]}}^{-1})+\mathbf{x}_{n-1|N}^H \mathbf{\mathbf{Q}^{[l]}}^{-1}\mathbf{x}_{n|N}}{\sum_{n = 2}^{N} \text{Tr}(\boldsymbol{\Sigma}_{n-1|N}{\mathbf{Q}^{[l]}}^{-1})+\mathbf{x}_{n-1|N}^H {\mathbf{Q}^{[l]}}^{-1}\mathbf{x}_{n-1|N}}$}
\end{align}
\vspace{-2mm}
and
\vspace{-1mm}
\begin{align}
\label{eq:m-step1}
& \resizebox{0.98\columnwidth}{!}{$\displaystyle \mathbf{Q}^{[l+1]} =\frac{1}{N}\sum_{n=1}^{N} \Big[\mathbf{x}_{n|N}\mathbf{x}_{n|N}^{H}\!+\!\boldsymbol{\Sigma}_{n|N}\!+\!{\alpha^{[l+1]}}^2 (\mathbf{x}_{n-1|N}\mathbf{x}_{n-1|N}^{H}\!+\!\boldsymbol{\Sigma}_{n-1|N})$} \nonumber \\[-4pt]
 &\qquad \qquad \quad\resizebox{.68\columnwidth}{!}{$\displaystyle\!\!\!\!\!\!\!\!\!-\!\alpha^{[l+1]} (\mathbf{x}_{n-1|N}\mathbf{x}_{n|N}^{H}\!+\!\mathbf{x}_{n|N}\mathbf{x}_{n-1|N}^{H}\!+\!2\boldsymbol{\Sigma}_{n,n-1|N}) \Big]$}.
\end{align}
\vspace{-5mm}

These iterations can be performed until convergence to a possibly local maximum. However, with even one such update, the overall algorithm forms a majorization-minimization (MM) procedure, generalizing the EM procedure and enjoying from similar convergence properties \cite{lange2004}. One possible implementation of this iterative procedure is described in Algorithm \ref{alg:DS3}. Once the {\sf DBMT} estimates of all the $K$ eigen-coefficients $\widehat{\mathbf{x}}_n^{(k)}$ are obtained, for $n = 1,2,\cdots ,N$ and $k = 1,2,\cdots,K$, the {\sf DBMT} spectrum estimate is constructed similar to (\ref{eq:mtsa}):
\begin{align}
\label{est:dbmt}
\widehat{D}_n(f_j) = \frac{1}{K}{\sum_{k=1}^{K}\left |\left(\widehat{\mathbf{x}}_{n}^{(k)}\right)_j\right|^2},
\end{align}
where $f_j := \frac{2\pi(j-1)}{J}$ for $j=1,2,\cdots, J$ and $n=1,2,\cdots,N$. {Confidence intervals can be computed by mapping the Gaussian confidence intervals for $\widehat{\mathbf{x}}_n^{(k)}$'s to the final {\sf DBMT} estimate.}

\subsection{The {\sf log-DBMT} Spectrum Estimation Algorithm}

We utilize a similar iterative procedure based on the EM algorithm to find the {\sf log-DBMT} spectrum estimate. As before, we treat $\mathbf{s}_n, n = 1,2,\cdots,N$ as hidden variables and $\theta, \mathbf{R}_{n},n = 1,2,\cdots,N$ as the unknown parameters to be estimated. In order to give more flexibility to the observation model, we consider the observation noise to be distributed as log-chi-square with degrees of freedom $2 \nu$, for some positive integer $\nu$ to be estimated. The density of each element of $\boldsymbol{\phi}_n^{(k)}$ is then given by:
\vspace{-3mm}
\begin{align}
p(\phi) = \frac{1}{2^{\nu} \Gamma(\nu)}\exp\left (\nu\phi-\frac{1}{2}\exp(\phi)\right).
\end{align} 
We can express the complete data log-likelihood as:
\begin{align}
\label{eq:complete-ll2}
&\resizebox{\columnwidth}{!}{$\displaystyle \log L(\nu, \theta, \mathbf{R}_{1:n})\!:=-\sum_{n = 1}^{N}\bigg[ \mathbf{1}_J^\top\left(\nu (\mathbf{s}_n - \boldsymbol{\psi}_n)+ \frac{1}{2}\exp(\boldsymbol{\psi}_n-\mathbf{s}_n)\right)$} \nonumber \\  
 \nonumber &+ J\bigg(\nu \log 2 + \log \Gamma(\nu)\bigg) + (\mathbf{s}_n\!-\theta\mathbf{s}_{n-1})^H \mathbf{R}_{n}^{-1}(\mathbf{s}_n\!-\theta\mathbf{s}_{n-1})\\
  &\qquad \qquad \qquad \qquad \quad \quad + \log |\det \mathbf{R}_n| \bigg] + {c},
\end{align}
{where $c$ represents the terms that do not depend on $\nu$, $\theta$, $(\mathbf{R}_{n})_{n=1}^N$ or $(\mathbf{s}_{n})_{n=1}^N$}.
Again assuming $\mathbf{R}_n = \mathbf{R}$ for all $n = 1,2, \cdots, N$ for simplicity, the following EM algorithm can be constructed:
\subsubsection{E-step}
Computation of the conditional expectation of the log-likelihood in (\ref{eq:complete-ll2}) requires evaluating $\mathbbm{E}[\mathbf{s}_n|\boldsymbol{\psi}_{1:N},\mathbf{R}^{[l]},\theta^{[l]},\nu^{[l]}]$ and $\mathbbm{E}[\exp(-\mathbf{s}_n)|\boldsymbol{\psi}_{1:N},\mathbf{R}^{[l]},\theta^{[l]},\nu^{[l]}]$ for $n = 1,2,\cdots,N$. Unlike the {\sf DBMT} estimation problem, the forward model in this case is non-Gaussian, and hence we cannot apply the Kalman filter and FIS to find the state expectations. To compute the conditional expectation, the distribution of $\mathbf{s}_n|\boldsymbol{\psi}_{1:n},\mathbf{R}^{[l]},\theta^{[l]},\nu^{[l]}$ or its samples are required \cite{Kitagawa1987}. Computation of the distribution $\mathbf{s}_n|\boldsymbol{\psi}_{1:n},\mathbf{R}^{[l]},\theta^{[l]},\nu^{[l]}$ involves intractable integrals and sampling from the distribution using numerical methods such as Metropolis-Hastings is not computationally efficient, especially for long data, given that it has to be carried out at every iteration. 
\begin{algorithm}[!t]
	\caption{\small The {\sf log-DBMT} Estimate of the $k$th log-Eigen-spectra}
	\label{alg:DS4}
	\begin{algorithmic}[1]
		\State Initialize: observations $\boldsymbol{\psi}_{1:N}^{(k)}$; initial guess ${\mathbf{s}^{[0]}_{1:N}}$; initial guess $\mathbf{R}^{[0]}$; initial conditions $\boldsymbol{\Omega}_{0|0}$; tolerance ${\sf tol} \in (0,10^{-3})$, Maximum Number of iteration $L_{\max} \in \mathbbm{N}^+$.
		\Repeat
		\State $l =0$ .
		\State Forward filter for $n=1,2,\cdots, N$:
		\vspace{-1.5mm}
		\begin{small}
		\begin{align*}
		&\mathbf{s}_{n|n-1} = \theta^{[l]} \mathbf{s}_{n-1|n-1}\\
		&\mathbf{\Omega}_{n|n-1} = {\theta^{[l]}}^2\mathbf{\Omega}_{n-1|n-1} + \mathbf{R}^{[l]}\\
		&\mathbf{s}_{n|n} = \mathbf{s}_{n|n-1} + \mathbf{\Omega}_{n|n-1}\left[\frac{1}{2}\exp(\boldsymbol{\psi}_n - \mathbf{s}_{n|n}) - \nu^{[l]} \mathbf{1}_J \right]\\
		&\mathbf{\Omega}_{n|n} = \mathbf{\Omega}_{n|n-1}^{-1} - \frac{1}{2}\text{diag} \{ \exp(\boldsymbol{\psi}_n - \mathbf{s}_{n|n}) \}
		\end{align*} 
		\end{small}
		\vspace{-5.5mm}
		\State \resizebox{.78\columnwidth}{!}{Backward smoother for $n=N-1,N-2,\cdots,1$}:
		\vspace{-1.5mm}
		\begin{small}
		\begin{align*}
		&\mathbf{A}_n = \theta^{[l]}\mathbf{\Omega}_{n|n}\mathbf{\Omega}_{n+1|n}^{-1}\\
		&\mathbf{s}_{n|N} = \mathbf{s}_{n|n} + \mathbf{A}_n(\mathbf{s}_{n+1|N} - \mathbf{s}_{n+1|n})\\
		&\mathbf{\Omega}_{n|N} = \mathbf{\Omega}_{n|n} + \mathbf{A}_n(\mathbf{\Omega}_{n+1|N}-\mathbf{\Omega}_{n+1|n})\mathbf{A}_n^H
		\end{align*}
		\end{small}
		\State \resizebox{.82\columnwidth}{!}{Covariance smoothing for $n=N-1,N-2,\cdots,1$}:
		\vspace{-1.5mm}
		\begin{small}
		\begin{align*}
		\hspace{-25.5mm}\mathbf{\Omega}_{n,n-1|N} = \mathbf{A}_{n-1}\mathbf{\Omega}_{n|N}
		\end{align*}
		\end{small}
		\State Let {\small $\widehat{\mathbf{S}}^{[l]} := [\mathbf{s}_{1|N}^{H},\mathbf{s}_{2|N}^{H},\cdots,\mathbf{s}_{N|N}^{H}]^{H}$}.
		\State Update $\nu^{[l+1]}$, $\theta^{[l+1]}$ and $\mathbf{R}^{[l+1]}$ as:
		\begin{small}
			\begin{align*}
			\resizebox{.9\columnwidth}{!}{$\displaystyle \theta^{[l+1]} = \frac{\sum_{n = 2}^{N} \text{Tr} (\boldsymbol{\Omega}_{n,n-1|N}{\mathbf{R}^{[l]}}^{-1})+\mathbf{s}_{n-1|N}^H {\mathbf{R}^{[l]}}^{-1}\mathbf{s}_{n|N}}{\sum_{n = 2}^{N} \text{Tr}(\boldsymbol{\Omega}_{n-1|N}{\mathbf{R}^{[l]}}^{-1})+\mathbf{s}_{n-1|N}^H {\mathbf{R}^{[l]}}^{-1}\mathbf{s}_{n-1|N}}$},
			\end{align*}
		\end{small}
		\vspace{-4mm}
		\begin{small}
			\begin{align*}
			\nu^{[l+1]} = \frac{1 - \log2 + \frac{1}{JN}\sum_{n = 1}^{N}\mathbf{1}_J^{\top}(\boldsymbol{\psi}_n^{[l]}\!-\!\mathbf{s}_n^{[l]}) - \digamma(\nu^{[l+1]})}{\frac{2}{JN}\sum_{n = 1}^{N}\mathbf{1}_J^{\top}\exp(\boldsymbol{\psi}_n^{[l]}\!-\!\mathbf{s}_n^{[l]})}, 
			\end{align*}
		\end{small}
		\vspace{-5mm}
		\begin{small}
			\begin{align*}
			&\resizebox{.9\columnwidth}{!}{$\displaystyle \mathbf{R}^{[l+1]}\!=\frac{1}{N}\sum_{n=1}^{N} \big[\mathbf{s}_{n|N}\mathbf{s}_{n|N}^{H}\!+\!\boldsymbol{\Omega}_{n|N}\!+\!{\theta^{[l+1]}}^2 (\mathbf{s}_{n-1|N}\mathbf{s}_{n-1|N}^{H}\!+\!\boldsymbol{\Omega}_{n-1|N})$} \nonumber \\
			&\qquad \qquad \qquad \resizebox{.7\columnwidth}{!}{$\displaystyle -\theta^{[l+1]} (\mathbf{s}_{n-1|N}\mathbf{s}_{n|N}^{H}\!+\!\mathbf{s}_{n|N}\mathbf{s}_{n-1|N}^{H}\!+\!2\boldsymbol{\Omega}_{n,n-1|N}) \big]$}.   
			\end{align*}
		\end{small}
		\vspace{-6mm}
		\State Set $l \leftarrow l+1$.
		\vspace{1mm}
		\Until{$\frac{\|\widehat{\mathbf{S}}^{[l]}-\widehat{\mathbf{S}}^{(l-1)}\|_2}{\|\widehat{\mathbf{S}}^{[l]}\|_2} < {\sf tol}$ or $l = L_{\max}$.}
		\State Output: Denoised log-eigen-spectra {$\widehat{\mathbf{S}}^{[L]}$} where $L$ is the index of the last iteration of the algorithm, and error covariance matrices $\boldsymbol{\Omega}_{n|N}$ for $n = 1,2,\cdots,N$ in from last iteration of the algorithm.
	\end{algorithmic}
\end{algorithm}
Since the posterior distribution is unimodal and a deviation from the Gaussian posterior, we approximate the distribution of $\mathbf{s}_n|\boldsymbol{\psi}_{1:n},\mathbf{R}^{[l]},\theta^{[l]},\nu^{[l]}$ as a Gaussian distribution by matching its mean and covariance matrix to the log-posterior in Eq. (\ref{eq:complete-ll2}). To this end, the mean is approximated by the mode of $f_{\mathbf{s}_n|\boldsymbol{\psi}_{1:n},\mathbf{R}^{[l]},\theta^{[l]},\nu^{[l]}}$ and the covariance is set to the inverse of the negative Hessian of the log-likelihood in (\ref{eq:complete-ll2}) \cite{fahrmeir1992,ssm2003}. Under this {approximation}, computing $\mathbbm{E}[\exp(-\mathbf{s}_n)|\boldsymbol{\psi}_{1:n},\mathbf{R}^{[l]},\theta^{[l]},\nu^{[l]}]$ is also facilitated thanks to the closed-form moment generating function of $\mathbf{z} \sim \mathcal{N}(\boldsymbol{\mu}, \boldsymbol{\Sigma})$:
\begin{align}
\label{eq:MGF}
\mathbbm{E}\left[\exp\left(\mathbf{a}^\top \mathbf{z}\right)\right] = \exp\left(\mathbf{a}^\top \boldsymbol{\mu} + \frac{1}{2}\mathbf{a}^\top \boldsymbol{\Sigma} \mathbf{a}\right).
\end{align}
Similar to the case of {\sf DBMT}, we can exploit the block tri-diagonal structure of the Hessian in (\ref{opt:log-DBMT}) to carry out the E-step efficiently using forward filtering and backward smoothing. 

\subsubsection{M-step} 
Once the conditional expectation of the log-likelihood in (\ref{eq:complete-ll2}) given $\boldsymbol{\psi}_{1:n}$, $\mathbf{R}^{[l]}$, $\theta^{[l]}$, $\nu^{[l]}$ is available, we can update $\mathbf{R}^{[l+1]}$ and $\theta^{[l+1]}$ using similar closed form equations as in (\ref{eq:m-step1}). But updating $\nu^{[l+1]}$  by maximizing the conditional expectation of the log likelihood in (\ref{eq:complete-ll2}) wrt. $\nu^{[l+1]}$ requires solving following nonlinear equation:
\begin{align}
\resizebox{.89\columnwidth}{!}{$\displaystyle \nu^{[l+1]} = \frac{1 - \log2 + \frac{1}{JN}\sum_{n = 1}^{N}\mathbf{1}_J^{\top}(\boldsymbol{\psi}_n^{[l]}\!-\!\mathbf{s}_n^{[l]}) - \digamma(\nu^{[l+1]})}{\frac{2}{JN}\sum_{n = 1}^{N}\mathbf{1}_J^{\top}\exp(\boldsymbol{\psi}_n^{[l]}\!-\!\mathbf{s}_n^{[l]})},$}
\end{align}
where $\digamma(\cdot)$ is the digamma function. We can use Newton's method to solve this equation up to a given precision. An implementation of the {\sf log-DBMT} is given by Algorithm \ref{alg:DS4}.
%
Note that unlike the {\sf DBMT} algorithm which pertains to a Gaussian observation model, the forward filtering step to compute $\mathbf{s}_{n|n}$ is nonlinear, and standard techniques such as Newton's method can be used to solve for $\mathbf{s}_{n|n}$. We use the {\sf log-DBMT} algorithm to find all the $K$ estimates of true log-spectra and construct the {\sf log-DBMT} estimate as:
\begin{align}
\label{est:logdbmt}
\widehat{D}_n(f_j) = \frac{1}{K} \sum_{k=1}^{K} \exp\left ( \left(\widehat{\mathbf{s}}^{(k)}_n\right)_j \right),
\end{align}
where $f_j := \frac{2\pi(j-1)}{J}$ for $j=1,2,\cdots, J$ and $n=1,2,\cdots,N$. Again, confidence intervals can be computed by mapping the Gaussian confidence intervals for $\widehat{\mathbf{s}}_n^{(k)}$'s to the final {\sf log-DBMT} estimate.

\subsection{Parameter Selection}
The window length $W$, design bandwidth $B$, and the number of tapers $K$ need to be carefully chosen. Since both proposed algorithms are motivated by the standard overlapping MT method, we use the same guidelines for choosing these parameters \cite{thomsonbc,reviewmtm}. The window length $W$ is determined based on the expected rate of change of the PSD (given domain-specific knowledge) in order to make sure that the quasi-stationarity assumption holds. The design bandwidth $B$ is chosen small enough to be able to resolve the dominant frequency components in the data, while being large enough to keep the time-bandwidth product $\rho := WB \geqslant 1$. The number of tapers $K$ is then chosen as $K \leqslant \lfloor2\rho\rfloor -1$ \cite{thomsonbc}.

\section{Application to Synthetic and Real Data} \label{sec:application}
Before presenting our theoretical analysis, we examine the performance of {\sf DBMT} and {\sf log-DBMT} spectrogram estimators on synthetic data, and then demonstrate their utility in two real world data applications, namely spectral analysis of human EEG during sleep and Electric Network Frequency signal detection.
\vspace*{-4.5mm}
\subsection{Application to Synthetic Data}
\label{sec:simulation}
\begin{figure}[!h]
\centering
\includegraphics[width=0.8\columnwidth]{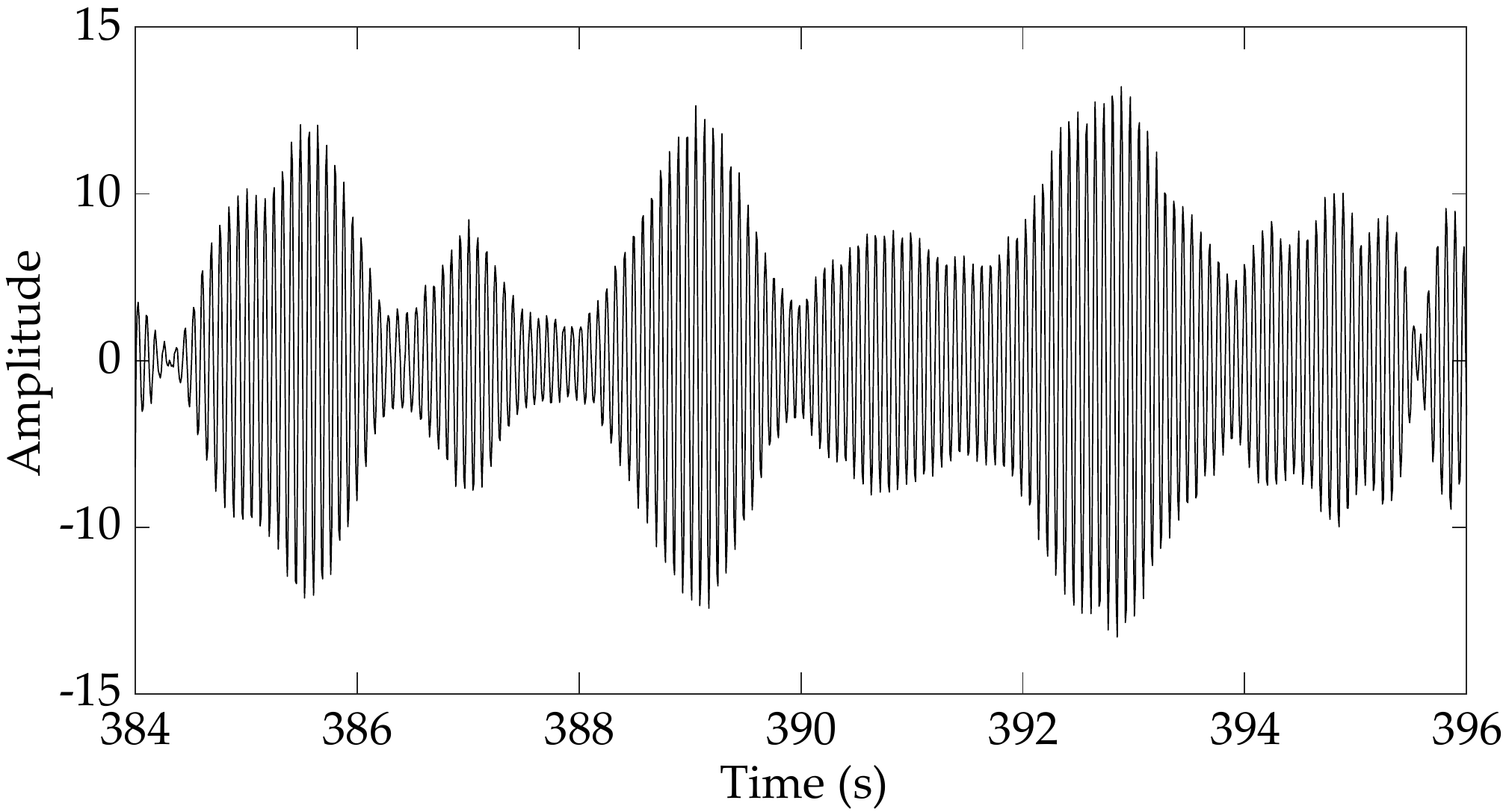}%
\vspace{-2mm}
\caption{\small Sample from the synthetic data from $t=384~\text{s}$ to $t=396~\text{s}$.}
\vspace{-4mm}
\label{fig:toy_example}
\end{figure}

The synthetic data consists of the linear combination of two amplitude-modulated and frequency-modulated processes with high dynamic range (i.e., high-Q). The amplitude-modulated component $y^{(1)}_t$ is generated through modulating an $\text{AR}(6)$ process tuned around $11~\text{Hz}$ by a cosine at a low frequency $f_0 = 0.02~\text{Hz}$. The frequency-modulated component $y^{(2)}_t$ is a realization of an $\text{ARMA}(6,4)$ with varying pole loci. To this end, the process has a pair of 3rd order poles at $\omega_t := 2 \pi f_t$ and $-\omega_t$, where $f_t$ increases from $5~\text{Hz}$, starting at $t=0$, every $\sim 26~\text{s}$ by increments of $0.48~\text{Hz}$, to achieve frequency modulation. In summary, the noisy observations are given by:
\begin{align}
y_t = y_t^{(1)} \cos(2\pi f_0t) + y_t^{(2)} + \sigma v_t,
\end{align} 
where $v_t$ is a white Gaussian noise process and $\sigma$ is chosen to achieve an SNR of $30~\text{dB}$. The process is truncated at $600~\text{s}$ to be used for spectrogram analysis. Figure \ref{fig:toy_example} shows a $12$ second sample window of the process.
\begin{figure*}[!t]
	\centering
	\includegraphics[width=0.8\textwidth]{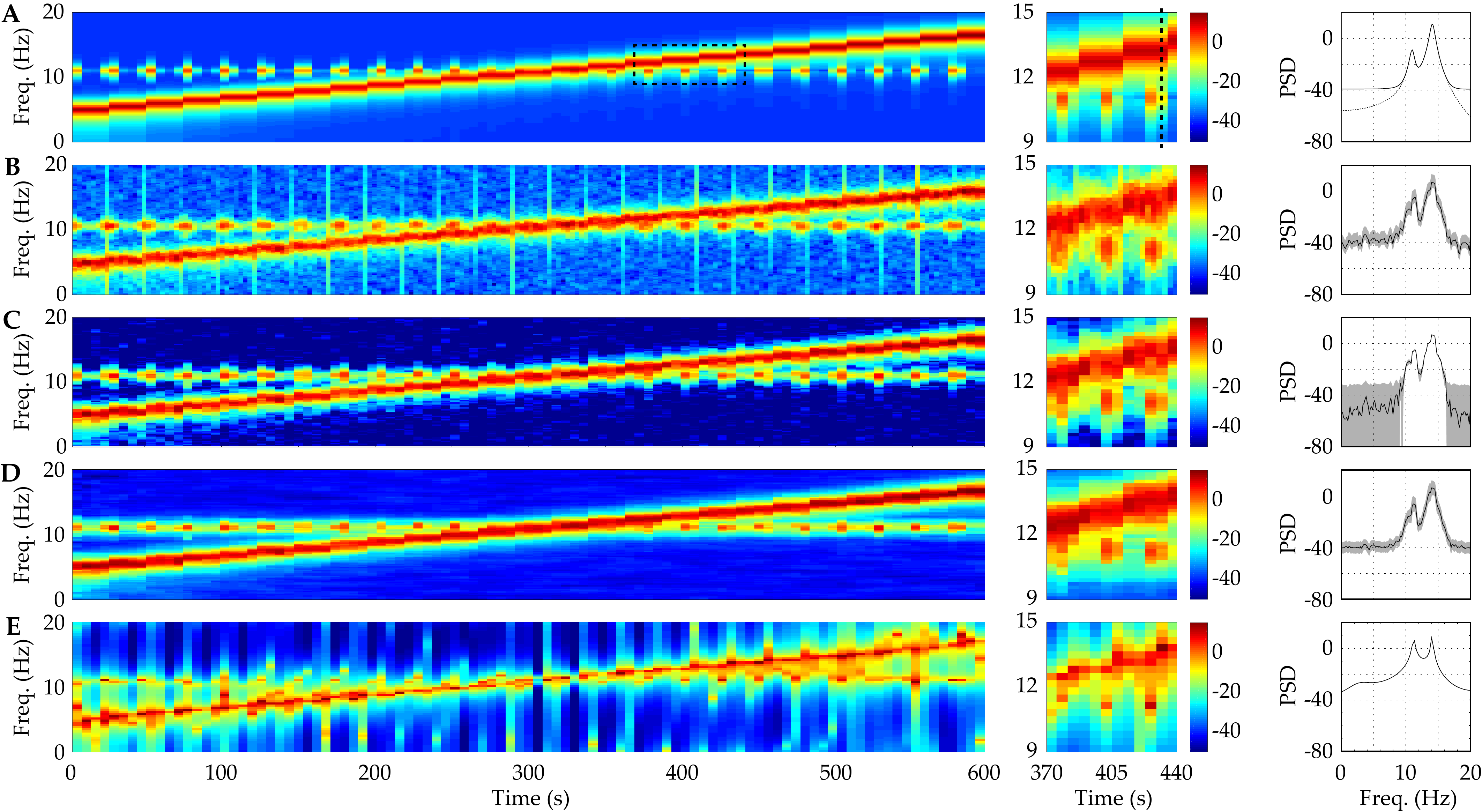}%
	\vspace{-2.0mm}
	\caption{\small Spectrogram analysis of the synthetic data. (A) Ground truth, (B) overlapping MT estimates, (C) {\sf DBMT} estimates, (D) {\sf log-DBMT} estimates, and (E) TFAR estimates. Left: spectrograms. Middle: zoomed-in views from $t=370~\text{s}$ to $t=440~\text{s}$. The color scale is in decibels. Right: PSDs corresponding to a window of length $6~\text{s}$ starting at $t=474~\text{s}$. Dashed and solid lines in row A show respectively the noiseless and noisy PSDs. Grey hulls show $95\%$ confidence intervals.}\label{fig:simulation}
	\vspace{-6mm}
\end{figure*}

{In addition to the standard overlapping MT, we present comparison to the TFAR method as an example of parametric state-space modeling approaches to non-stationary spectral analysis. This method is known to be well suited to processes whose time-varying spectra exhibits sharp peaks, i.e., signals consisting of several narrow-band components \cite{Jachan2007TFARMA}. The TFAR model is defined by the input-output relation
\begin{align*}
y_t\!:=\!-\!\sum_{m=1}^{M_A}\sum_{l=-L_A}^{L_A} \!\!a_{m,l}e^{i\frac{2\pi}{N}lt} y_{t-m}\!+ \!\sum_{l=-L_B}^{L_B}\!\!b_{0,l}e^{i\frac{2\pi}{N}lt}e_t,
\end{align*}
where $e_t$ is a stationary white noise process with unit variance, and $(a_{m,l})_{m=1,l=-L_A}^{M_A,L_A}$ and $(b_{0,l})_{l=-L_B}^{L_B}$ are the autoregressive (AR) and zero-delay moving average (MA) parameters,  respectively. The integers $M_A$ and $L_A$ are respectively the delay and Doppler model orders of the AR component and $L_B$ denotes the Doppler model order of the zero-delay MA component. The AR and MA parameters are estimated by solving the time-frequency Yule-Walker equations, from which the evolutionary spectra can be constructed (See the methods described in \cite{Jachan2007TFARMA} for more details).} \newline
\indent Figure \ref{fig:simulation} shows the true as well as estimated spectrograms by the standard overlapping MT, {\sf DBMT}, {\sf log-DBMT} and the TFAR estimators. Each row consists of three panels: the left panel shows the entire spectrogram; the middle panel shows a zoomed-in spectrotemporal region marked by the dashed box in the left panel; and the right panel shows the PSD along with confidence interval (CI) in gray hull, at a selected time point marked by a dashed vertical line in the middle panel. {Note that for the standard MT estimates, the CIs are constructed assuming a $\chi_{2K}^2$ distribution of the estimates around the true values \cite{thomsonbc,percival1993}, whereas for {\sf DBMT} and {\sf log-DBMT} estimate by mapping the Gaussian confidence intervals for eigen-coefficients or eigen-spectra to the final estimates.} {We were not able to evaluate the CIs for the TFAR estimates, since to the best of our knowledge we are not aware of any method to do so.} Figure \ref{fig:simulation}\textit{A} shows true spectrogram of the synthetic process, in which the existence of both amplitude and frequency modulations makes the spectrogram estimation a challenging problem. \newline  

\indent Fig. \ref{fig:simulation}\textit{B} shows the standard overlapping MT spectrogram estimate. We used windows of length $6~\text{s}$ and the first $3$ tapers corresponding to a time-bandwidth product of $3$ and $50\%$ overlap to compute the estimates (note that the same window length, tapers and time-bandwidth product are used for the {\sf DBMT} and {\sf log-DBMT} estimators). Although the standard MT spectrogram captures the dynamic evolution of both components, it is blurred by the background noise and picks up spectral artifacts (i.e., vertical lines) due to window overlap, frequency mixing, and sampling noise. Fig. \ref{fig:simulation}\textit{C} demonstrates how the {\sf DBMT} spectrogram estimate overcomes these deficiencies of the overlapping MT spectrogram: the spectrotemporal localization is sharper and smoother across time, artifacts due to overlapping between windows are vanished, and frequency mixing is further mitigated. By comparing the right panel of the second and third rows, two important observations can be made: first, the {\sf DBMT} captures the true dynamic range of the original noiseless PSD, while the standard MT estimate fails to do so. Second, the CIs in Fig. \ref{fig:simulation}\textit{C} as compared to \ref{fig:simulation}\textit{B} are wider when the signal is weak (e.g., near $5~\text{Hz}$) and tighter when the signal is strong (e.g., near $11~\text{Hz}$). The latter observation highlights the importance of the model-based confidence intervals in interpreting the denoised estimates of {\sf DBMT}: while the most likely estimate (i.e., the mean) captures the true dynamic range of the noiseless PSD, the estimator does not preclude cases in which the noise floor of $-40~\text{dB}$ is part of the true signal, while showing high confidence in detecting the spectral content of the true signal that abides by the modeled dynamics. \newline
\indent Next, Fig. \ref{fig:simulation}\textit{D} shows the {\sf log-DBMT} spectrogram estimate, which shares the artifact rejection feature of the {\sf DBMT} spectrogram. However, the {\sf log-DBMT} estimate is smoother than both the {standard overlapping MT} and {\sf DBMT} spectrograms in time as well as in frequency (see the zoomed-in middle panels), due to its sampling noise mitigation feature (by design). In addition, similar to the standard {MT} estimate, the {\sf log-DBMT} estimator is not able to denoise the estimate by removing the observation noise. Though, the confidence intervals of the {\sf log-DBMT} PSD estimate are tighter than those of the standard overlapping MT estimate due to averaging across multiple windows via Bayesian filtering/smoothing. As we will show in Section \ref{sec:theory}, these qualitative observations can be established by our theoretical analysis.

{Finally, Fig. \ref{fig:simulation}\textit{E} shows the TFAR spectral estimate. The model orders are chosen as $M_A =20$, $L_A = 25$, $L_B = 25$, large enough to allow the parametric model to achieve high time-frequency resolution. As it is shown in the right panel, the TFAR method provides the smoothest estimate along the frequency axis. However, it is not successful in capturing the true dynamic range of the signal due to spectral leakage (See middle and right panels). In addition, it is contaminated by similar vertical frequency artifacts as in the case of the MT spectrogram (See Figure \ref{fig:simulation}\textit{A}).} \newline
\indent In the spirit of easing reproducibility, we have deposited a MATLAB implementation of these algorithms on the open source repository GitHub \cite{code}, which generates Figure \ref{fig:simulation}.
\subsection{Application to EEG data}

\begin{figure*}[!t]
	\centering
	\includegraphics[width=0.80\textwidth]{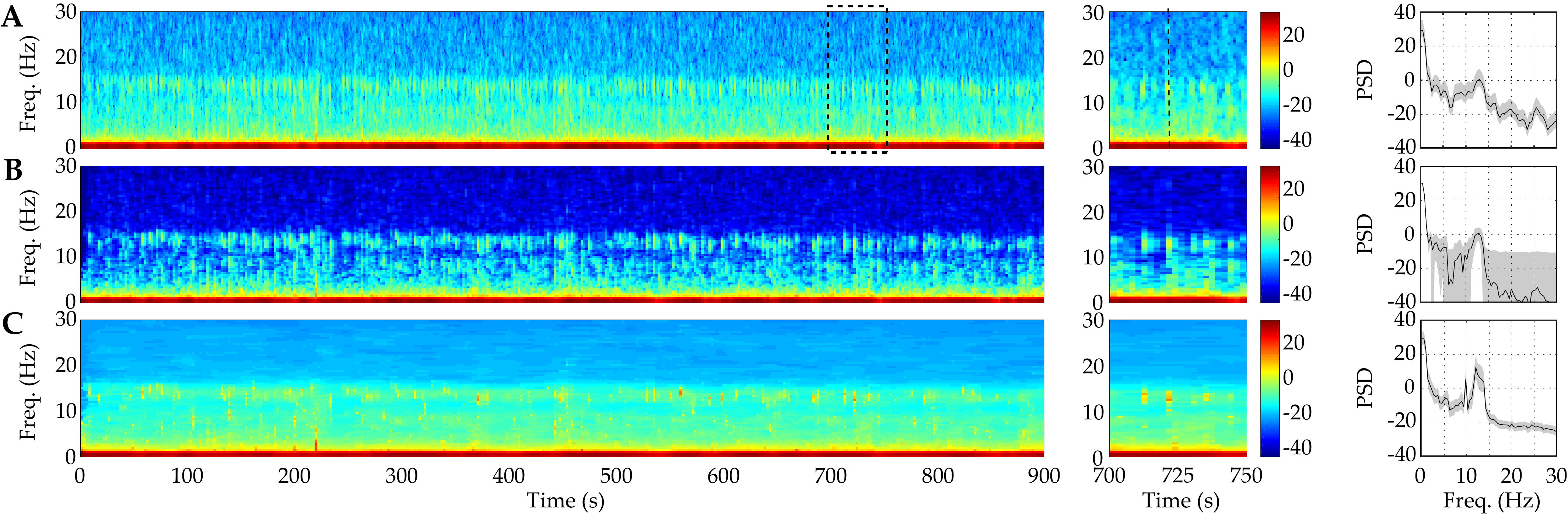}%
	\caption{\small Spectrogram analysis of the EEG data. (A) overlapping MT estimates, (B) {\sf DBMT} estimates, and (C) {\sf log-DBMT} estimates. Left: spectrograms. Middle: zoomed-in views from $t=700~\text{s}$ to $t=750~\text{s}$. The color scale is in decibels. Right: PSD estimate corresponding to a window of length $2.25~\text{s}$ starting at $t=722.25~\text{s}$. Grey hulls show $95\%$ confidence intervals.}
	\label{fig:sleepdata}
	\vspace*{-5mm}
\end{figure*}
\begin{figure*}[!b]
	\centering
	\vspace*{-4mm}
	\includegraphics[width=0.80\textwidth]{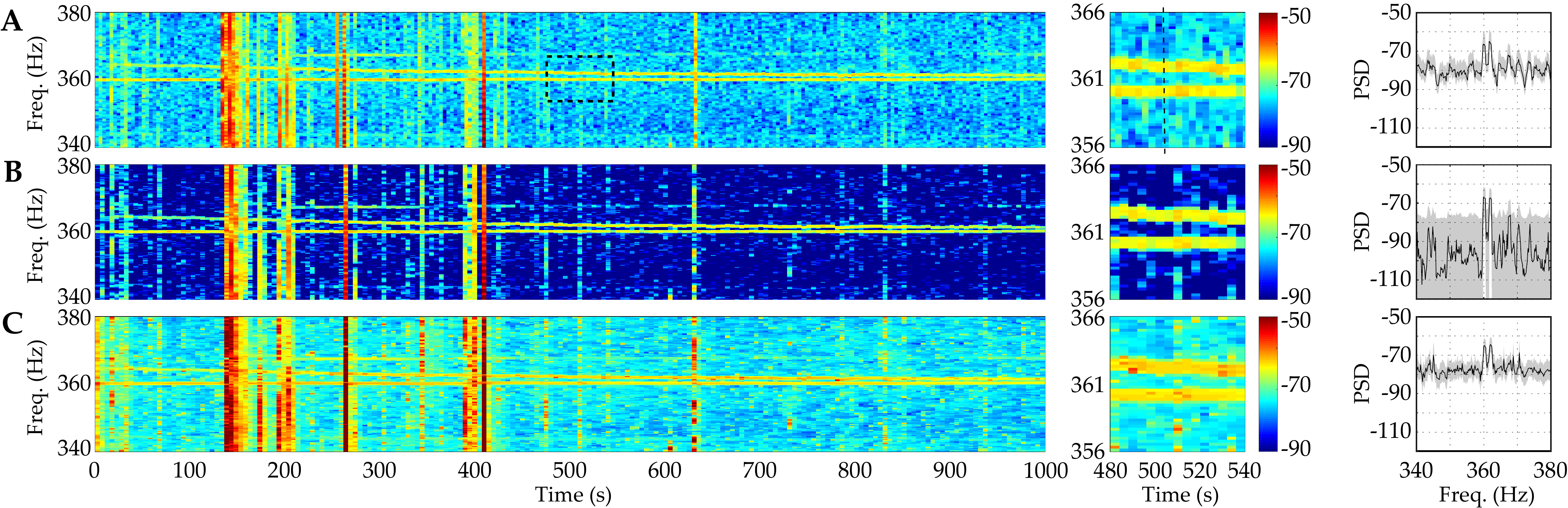}%
   \vspace*{-1.5mm}	
	\caption{\small Spectrogram analysis of the ENF data. (A) overlapping MT estimates, (B) {\sf DBMT} estimates, and (C) {\sf log-DBMT} estimates. Left: spectrograms. Middle: zoomed-in views from $t=480~\text{s}$ to $t=540~\text{s}$. The color scale is in decibels. Right: PSD estimate corresponding to a window of length $5~\text{s}$ starting at $t=505~\text{s}$. Grey hulls show $95\%$ confidence intervals.}
	\label{fig:ENFdata}
\end{figure*}
To illustrate the utility of our proposed spectrogram estimators, we apply them to human EEG data recorded during sleep. {In the interest of space, in the remainder of this section, we only present comparisons with the MT spectrogram as a non-parametric benchmark.} The EEG data set is available online as part of the SHHS Polysomnography Database (\url{https://www.physionet.org/pn3/shhpsgdb/}). The data is $900~\text{s}$ long during stage 2 sleep, and sampled at $250~\text{Hz}$. During stage 2 sleep, the EEG is known to manifest delta waves ($0-4~\text{Hz}$) and sleep spindles (transient wave packets with frequency $12-14~\text{Hz}$) \cite{spindles2003, delta1933}. Accurate localization of these spectrotemporal features has significant applications in studying sleep disorders and cognitive function \cite{spindles2003}. Since the transient spindles occur at a time scale of seconds, we choose a window length of $2.25~\text{s}$ for all algorithms (with $50\%$ overlap for the standard overlapping MT estimate). We also chose a time-bandwidth product of $2.25$ for all algorithms, in order to keep the frequency resolution at $2~\text{Hz}$. Figs. \ref{fig:sleepdata}\textit{A}, \textit{B} and \textit{C} show the MT, {\sf DBMT} and {\sf log-DBMT} spectrogram estimates, respectively, with a similar presentational structure as in Fig. \ref{fig:simulation}. As the middle panels reveal, the overlapping MT estimate is not {able} to clearly distinguish the delta waves and sleep spindles due to high background noise. The {\sf DBMT} estimate shown in Fig. \ref{fig:sleepdata}\textit{B}, however, provides a significantly denoised spectrogram, in which the delta waves and sleep spindles are visually separable. The {\sf log-DBMT} estimator shown in Fig. \ref{fig:sleepdata}\textit{C} provides significant spectrotemporal smoothing, and despite not fully reducing the background noise, provides a clear separation of the delta waves and spindles (see the PSD in the right panel). Similar to the analysis of synthetic data, the same observations regarding the confidence intervals of the estimators can be made.

\vspace{-2mm}
\subsection{Application to ENF data}

Finally, we examine the performance of our proposed algorithms in tracking the Electrical Network Frequency (ENF) signals from audio recordings. The ENF signal corresponds to the supply frequency of the power distribution network which is embedded in audio recordings \cite{garg2013seeing,Wu2013}. {The instantaneous values of this time-varying frequency and its harmonics form the ENF signal.} The ability to detect and track the spectrotemporal dynamics of ENF signals embedded in audio recordings has shown to be crucial in data forensics applications \cite{garg2013seeing}. 

Fig. \ref{fig:ENFdata}\textit{A} shows the spectrogram estimates around the sixth harmonic of the nominal $60~\text{Hz}$ ENF signal (data from \cite{Wu2013}). We used $1000~\text{s}$ of audio recordings, and constructed spectrograms with windows of length $5~\text{s}$ and using the first $3$ tapers corresponding to a time-bandwidth product of $3$ for all three methods (with $25\%$ overlap for the overlapping MT estimate). The two dominant components around the sixth ENF harmonic exhibit temporal dynamics, but are hard to distinguish from the noisy background. Fig. \ref{fig:ENFdata}\textit{B} shows the {\sf DBMT} spectrogram, in which the background noise is significantly suppressed, yielding a crisp and temporally smooth estimate of the ENF dynamics. The {\sf log-DBMT} estimate is shown in Fig. \ref{fig:ENFdata}\textit{C}, which provides higher spectrotemporal smoothness than the standard MT estimate. Although the {\sf log-DBMT} shows smaller variability in the estimates (middle and right panels), the gain is not as striking as in the cases of synthetic data and EEG analysis, due to the usage of longer windows which mitigates the sampling noise for all algorithms. Similar observations as in the previous two cases regarding the statistical confidence intervals can be made, which highlight the advantage of modeling the spectrotemporal dynamics in spectrogram estimation.

\section{Theoretical Analysis}\label{sec:theory}
\subsection{Filter Bank Interpretation}
In order to characterize the spectral properties of any non-parametric spectrum estimator, the tapers applied to the data need to be carefully inspected. In the MT framework, the dpss sequences are used as tapers, which are known to produce negligible side-lobes in the frequency domain \cite{mtm,percival1993}. The {\sf DBMT} and {\sf log-DBMT} algorithms also use the dpss tapers to alleviate the problem of frequency leakage. However, because of the stochastic continuity constraint we introduced, the estimate associated to any given window is now a function of the data in \emph{all} the windows. Therefore, the theoretical properties of the MT method do not readily apply to our estimators.

To characterize the statistical properties of our estimates, we first need take a detour from the usual analysis of spectrum estimation techniques. In what follows, we mainly focus on the {\sf DBMT} algorithm for the sake of presentation. By virtue of the FIS procedure under the assumptions that: 1) the window length $W$ is an integer multiple of $J$, the number of discrete frequencies, so that $\mathbf{F}_n = \mathbf{F}_1, \forall n$, and 2) the state noise covariance matrices are time-invariant, i.e., $\mathbf{Q}_n = \mathbf{Q}, \forall n$, one obtains the following expansion of $\mathbf{x}_{n|N}^{(k)}$ in terms of the observed data \cite{ba2014}:
\begin{align}
\label{FB3}
\mathbf{{x}}_{n|N}^{(k)} = \sum_{s = 1}^{n-1}\!\prod_{m = s}^{n-1}\!\big[\alpha(\mathbf{I}\!-\!\mathbf{K}_m&\mathbf{F}_m)\big] \mathbf{K}_s \mathbf{U}^{(k)} \widetilde{\mathbf{y}}_s + \mathbf{K}_n \mathbf{U}^{(k)} \widetilde{\mathbf{y}}_n \nonumber \\[-8pt]
+& \sum_{s = n+1}^{N}\!\prod_{m = n}^{s}\!\mathbf{B}_m \mathbf{K}_s \mathbf{U}^{(k)} \widetilde{\mathbf{y}}_s.
\end{align}
In other words, the {\sf DBMT} algorithm maps the entire data $\widetilde{\mathbf{y}} := [\widetilde{y}_1,\widetilde{y}_2,\cdots, \widetilde{y}_T]^\top$ to the {vector of coefficients} $\widehat{\mathbf{X}}^{(k)}$ according to \cite{ba2014}: 
\begin{equation}
\label{FilterBank2}
\widehat{\mathbf{X}}^{(k)} = \mathbf{G}^{(k)} \mathbf{F}^{H}\mathbf{U}^{(k)}\widetilde{\mathbf{y}},
\end{equation}
where $\mathbf{F}$ and $\mathbf{U}^{(k)}$ are block-diagonal matrices with $\mathbf{F}_1$ and $\mathbf{U}_{k} := \text{diag}[\mathbf{u}^{(k)}]$ as the diagonal blocks, respectively, and $\mathbf{G}$ is a weighting matrix which depends only on $\mathbf{Q}_\infty = \lim_{l\rightarrow \infty} \mathbf{Q}^{[l]}$, {$\alpha_\infty = \lim_{l \rightarrow \infty} \alpha^{[l]}$,} and window length, $W$. The rows of $\mathbf{G}^{(k)}\mathbf{F}^{H}\mathbf{U}^{(k)}$ form a filter bank whose output is equivalent to the time-frequency representation.

In order to continue our analysis, we make two assumptions common in the analysis of adaptive filters\cite{haykin1991,anderson1979optimal}. First, we assume that the parameter {estimates $\mathbf{Q}_\infty$ and $\alpha_\infty$ are close enough to the true values of $\mathbf{Q}$ and $\alpha$, and therefore replace them by $\mathbf{Q}$ and $\alpha$, i.e., as if the true parameters were known.} Note that we have discarded the dependence of $\mathbf{Q}$ and $\alpha$ on $k$ for lucidity of analysis. Second, noting that $\alpha(\mathbf{I}-\mathbf{K}_m\mathbf{F}_m) = \alpha\boldsymbol{\Sigma}_{m|m} \boldsymbol{\Sigma}_{m|m-1}^{-1}$ and $\mathbf{B}_m = \alpha\mathbf{\Sigma}_{m|m}\mathbf{\Sigma}_{m+1|m}^{-1}$ and that in steady state we have $\boldsymbol{\Sigma}_{m|m} := \boldsymbol{\Sigma}_\infty $ and $\boldsymbol{\Sigma}_{m|m-1} = \alpha^2\boldsymbol{\Sigma}_\infty + \mathbf{Q}$, Eq. (\ref{FB3}) can be approximated by:
\vspace{-2mm}
\begin{align}\label{eq:xn}
\mathbf{{x}}_{n|N}^{(k)} = \sum_{s=1}^{N}\mathbf{ \Lambda}^{|s-n|} \mathbf{\Gamma} \mathbf{F}_s^{H}\mathbf{U}^{(k)}\widetilde{\mathbf{y}}_s,
\end{align}
\noindent for $1 \ll n \ll N$, where $ \mathbf{\Lambda} = \alpha \mathbf{\Sigma}_\infty (\alpha^2 \mathbf{\Sigma}_\infty + \mathbf{Q})^{-1}$, and $\mathbf{\Gamma} = (\alpha^2 \mathbf{\Sigma}_\infty + \mathbf{Q})\big[\mathbf{I}-rW\big((\alpha^2 \mathbf{\Sigma}_\infty + \mathbf{Q})^{-1}+rW\mathbf{I}\big)^{-1}\big]$. That is, for values of $n$ far from the data boundaries, the weighting matrix is equivalent to a weighted set of dpss tapers in matrix form acting on all the data windows with an exponential decay with respect to the $n$th window. 
\begin{figure}[!t]
	\centering
	\includegraphics[width=0.99\columnwidth]{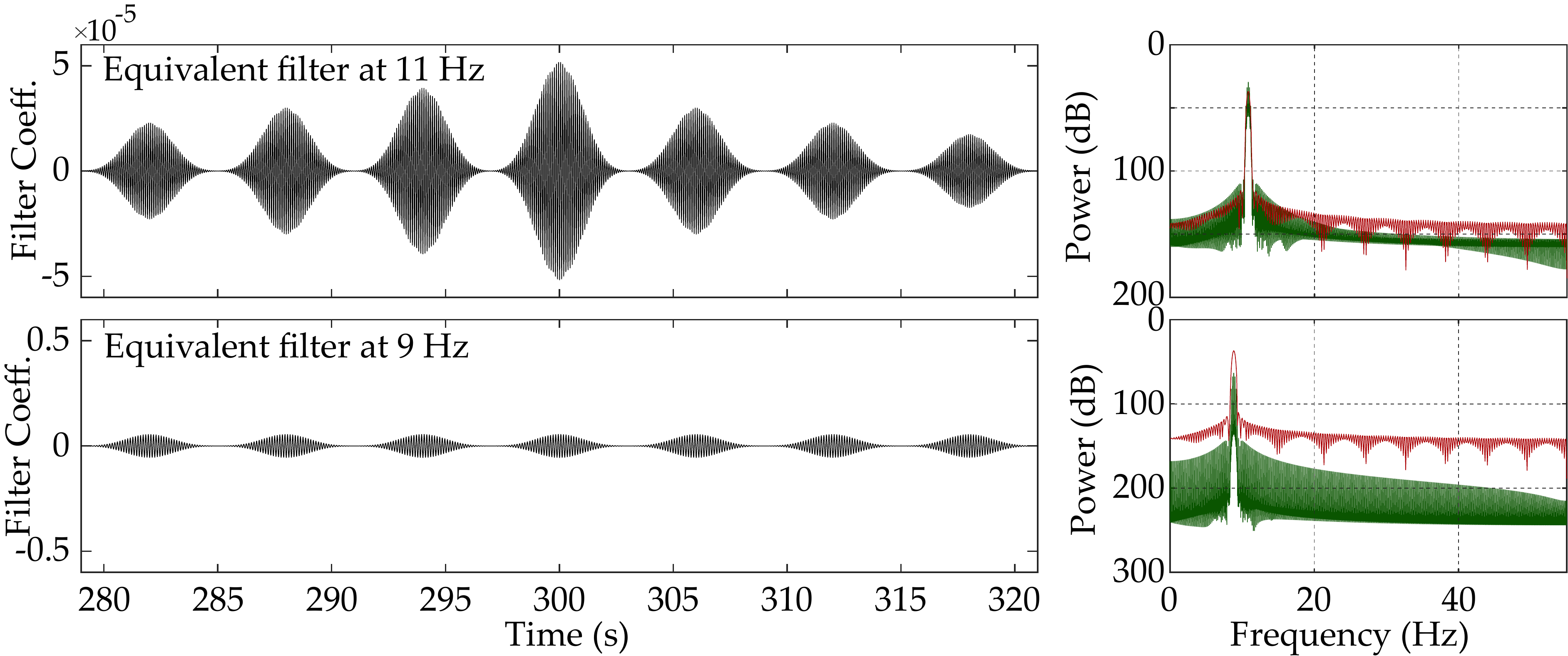}%
	\vspace{-1.5mm}
	\caption{\small {Equivalent filters corresponding to the first taper of the {\sf DBMT} estimate of the synthetic data example. Left: equivalent filters in time around $t=300~\text{s}$. Right: equivalent filters of MT (red) and {\sf DBMT} (green) in frequency.}}
	\vspace*{-5mm}
	\label{fig:filter}
\end{figure}
 {As an example, the equivalents filters of the {\sf DBMT} estimator corresponding to the first taper for the $11$Hz and $9$Hz frequencies around $300$ secs, from the synthetic data example are shown in Fig. \ref{fig:filter}. They are also compared to the equivalent filters corresponding to first taper of standard MT method in the frequency domain.} As apparent from Fig. \ref{fig:filter}, the weighting matrix sets the gain of these filters in an adaptive fashion across \emph{all} windows, unlike the standard MT method which only uses the data in window $n$. In addition, the filter corresponding to frequency of $9$Hz, which is negligible in the data, is highly attenuated, resulting in significant noise suppression. In this sense, the proposed estimation method can be identified as a \emph{data-driven denoising method} for constructing time-frequency representations given noisy time series data. Next, we will characterize the performance of the {\sf DBMT} estimator in terms of bias-variance trade-off.
\vspace{-4mm}
\subsection{Bias and Variance Analysis}

We first consider the implication of the stochastic continuity constraint of Eq. (\ref{eq:scc1}) on the evolution of the orthogonal increment processes governing the time series data. We first assume that the parameters $\alpha^{(k)} = \alpha$, for all $k=1,2,\cdots,K$. Suppose that the data in window $n$ has a Cram\'{e}r representation with an orthogonal increment process $dz_n(f)$, $n=1,2,\cdots,N$. Then, one way to achieve the stochastic continuity of Eq. (\ref{eq:scc1}) is to assume: \vspace{-2mm}
\begin{equation}\label{eq:z-ss}
d z_{n+1}(f) = \alpha d z_{n}(f) + d \epsilon_n (f),\vspace{-2mm}
\end{equation}
where $d \epsilon_n(f)$ is a Gaussian orthogonal increment process, independent of $d z_{n}(f)$. 
In the forthcoming analysis we also assume the locally stationarity condition, i.e., the generalized Fourier transform of the process remains stationary within each window. This assumption is common in the analysis of non-parametric spectral estimators \cite{reviewmtm,thomsonbc}. Finally, we assume a scaling of $K,N,W \rightarrow \infty$, $B \rightarrow 0$, $BW \rightarrow \rho$, for some constant $\rho$ \cite{lii2008}. The following theorems characterize the bias and variance of the {\sf DBMT} estimator:
\vspace*{-2mm}
\begin{thm}
\label{theorem1}
Suppose that the locally stationary process $y_t$ is governed by orthogonal increment processes evolving according to the dynamics $dz_{n+1}(f) = \alpha dz_n(f) + d \epsilon_n(f)$, with $\alpha < 1$, where the noise process $d \epsilon_n(f), \forall f \in (-1/2,1/2]$ is a zero-mean Gaussian increment process with variance $q(f)>0$, independent of $dz_n(f)$. If the process is corrupted by additive zero-mean Gaussian noise with variance $\sigma^2$, then {for $f \in \{f_1,f_2,\cdots,f_J\}$,} the {\sf DBMT} estimate satisfies:
\begin{align}
\nonumber & \resizebox{\columnwidth}{!}{$\displaystyle \left|\mathbbm{E}[\widehat{D}_{n}(f)] - D(f) \right| \leqslant \bigg(1-\frac{1}{K}\sum_{k=1}^{K}\lambda_k\bigg) \kappa_n(f) \sup_{f} \{D(f)\}$}\\
\nonumber & \qquad \qquad \qquad  \quad \! + |1-\kappa_n(f)| D(f) + \mu_n(f) \sigma^2 + \kappa_n(f) o(1),
\end{align}
where $\lambda_k$ is the eigenvalue associated with the $k$th PSWF, $D(f) :=  q(f)/(1-\alpha)$, and $\kappa_n(f), \mu_n(f)$ are functions of $\alpha$ and $q(f)$ and explicitly given in the proof.
\end{thm}
\vspace*{-3mm}
\begin{thm} 
\label{theorem2}
Under the assumptions of Theorem \ref{theorem1}, the variance of the {\sf DBMT} estimate $\widehat{D}_n(f)$ satisfies: 
\begin{align}
	\nonumber {\sf Var}\left \{\widehat{D}_{n}(f) \right \} \leqslant \frac{2}{K}  \left[\sup_{f} \{\kappa_n(f)D(f) + \mu_n(f)\sigma^2\}\right]^2.
\end{align}

\end{thm}

The proofs of Theorems \ref{theorem1} and \ref{theorem2} integrate the treatment of \cite{lii2008} with the structure of the FIS estimates, and are presented in Appendix \ref{A1}. In order to illustrate the implications of these theorems, several remarks are in order: 

\textit{\textbf{Remark 1.}}
The function $\kappa_n(f)$ controls the trade-off between bias and variance: for values of $\kappa_n(f) < 1$, the bound on the variance decreases while the bias bound increases, and for $\kappa_n(f) \approx 1$, all the terms in the bias bound become negligible, while the variance bound increases. The function $\mu_n(f)$, on the other hand, reflects observation noise suppression in both the bias and variance. Note that these upper bounds are tight and achieved for a signal with flat spectrum.
 
\textit{\textbf{Remark 2.}}
The bias and variance bounds of \cite{lii2008} for the standard MT method can be recovered by setting $\kappa_n(f) =1 $, $\mu_n(f) = 1$, and $\sigma^2 = 0$ in the results of Theorems \ref{theorem1} and \ref{theorem2}, i.e., in the absence of signal dynamics and {measurement} noise. For the {\sf DBMT} estimator, signal and {measurement} noise variances, respectively contribute to the bias/variance upper bounds in different fashions through $\kappa_n(f)$ and $\mu_n(f)$, due to the distinction of the signal and {measurement} noise in our state-space model. In contrast, in the standard MT method, possible {measurement} noise is treated in the same way as the true data, and hence both the signal and noise variances have equal contributions in the estimator bias/variance. 

\begin{figure}[t!]
\centering
\includegraphics[width=.88\columnwidth]{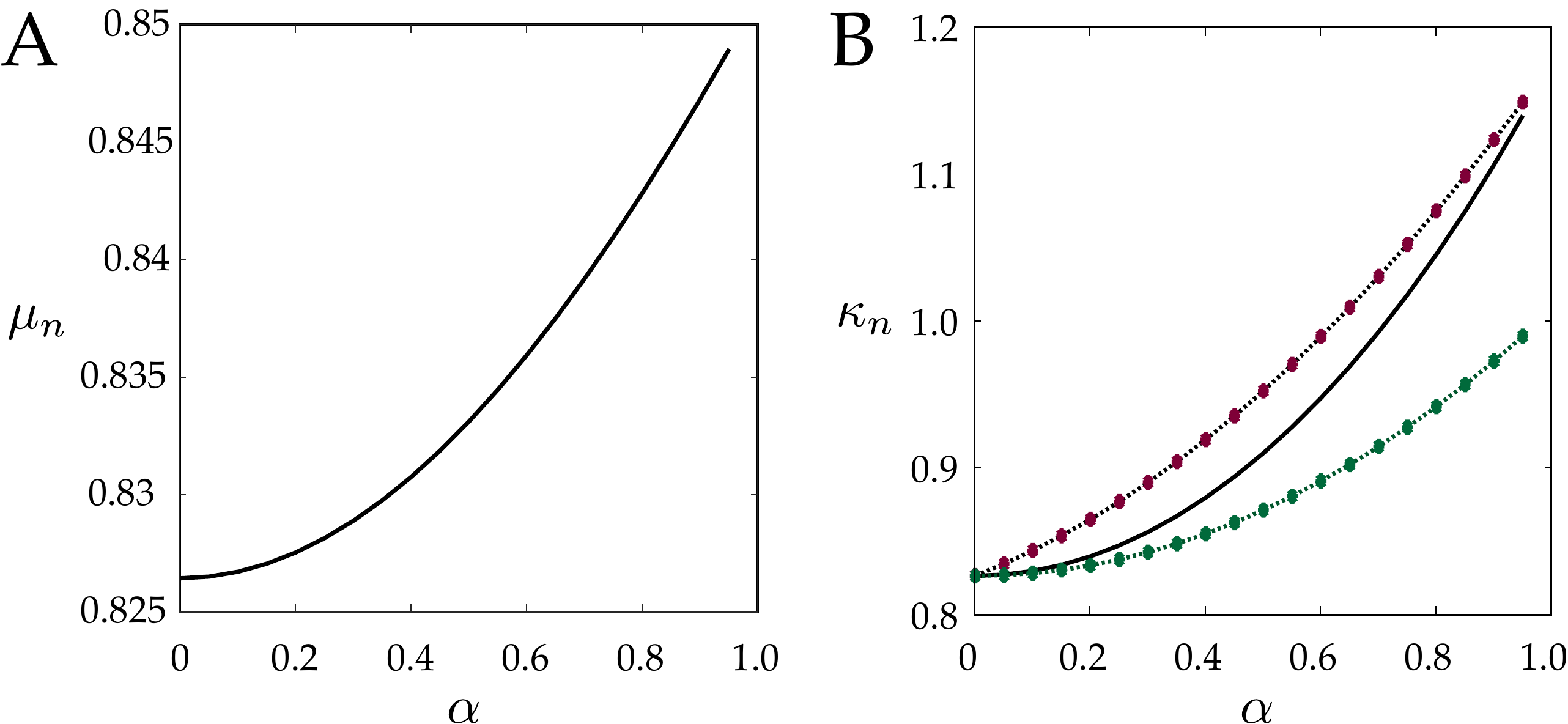}
\caption{\small (A) $\mu_n$ versus $\alpha$, (B) $\kappa_n$ and its upper/lower bounds versus $\alpha$ for $N = 100$, $n = 50$, and $q/\sigma^2 = 10$.}
\label{fig:kappa_vs_alpha}
\vspace{-6mm}
\end{figure}

The functions $\kappa_n(f)$ and $\mu_n(f)$ do not have closed-form expressions with respect to the state-space parameters $\alpha$, $\sigma^2$ and $q(f)$. In order to illustrate the roles of $\kappa_n(f)$ and $\mu_n(f)$, we consider the scenario under which the upper bounds on the bias and variance are achieved, i.e., $q(f)$ being independent of $f$, and hence $\kappa_n(f) = \kappa_n$ and $\mu_n(f) = \mu_n$, $\forall f$. In this scenario, even though the dependence of $\mu_n$ and $\kappa_n$ on the state-space parameters are quite involved, it is possible to obtain upper and lower bounds on $\kappa_n$ and $\mu_n$. As it is shown in Proposition \ref{prop:kappa} in Appendix \ref{A3}, the main parameters determining the behavior of $\kappa_n$ and $\mu_n$ are $q_n/\sigma$ (i.e., the SNR) and $\alpha$ (i.e., temporal signal dependence). Here, we present a numerical example for clarification. Fig. \ref{fig:kappa_vs_alpha}\textit{A} and \textit{B} shows the plot of $\mu_n$ vs. $\alpha$ and $\kappa_n$ vs. $\alpha$ for $n = 50$ and $q/\sigma^2 = 10$. It is apparent that $\mu_n$ increases with $\alpha$ and does not exceed $1$. The fact that $\mu_n < 1$ implies that the {\sf DBMT} estimator achieves a higher noise suppression compared to the standard MT method. This fact agrees with the noise suppression performances observed in Section \ref{sec:application}.

\begin{figure}[t!]
\centering
\vspace{0mm}
\includegraphics[width=.8\columnwidth]{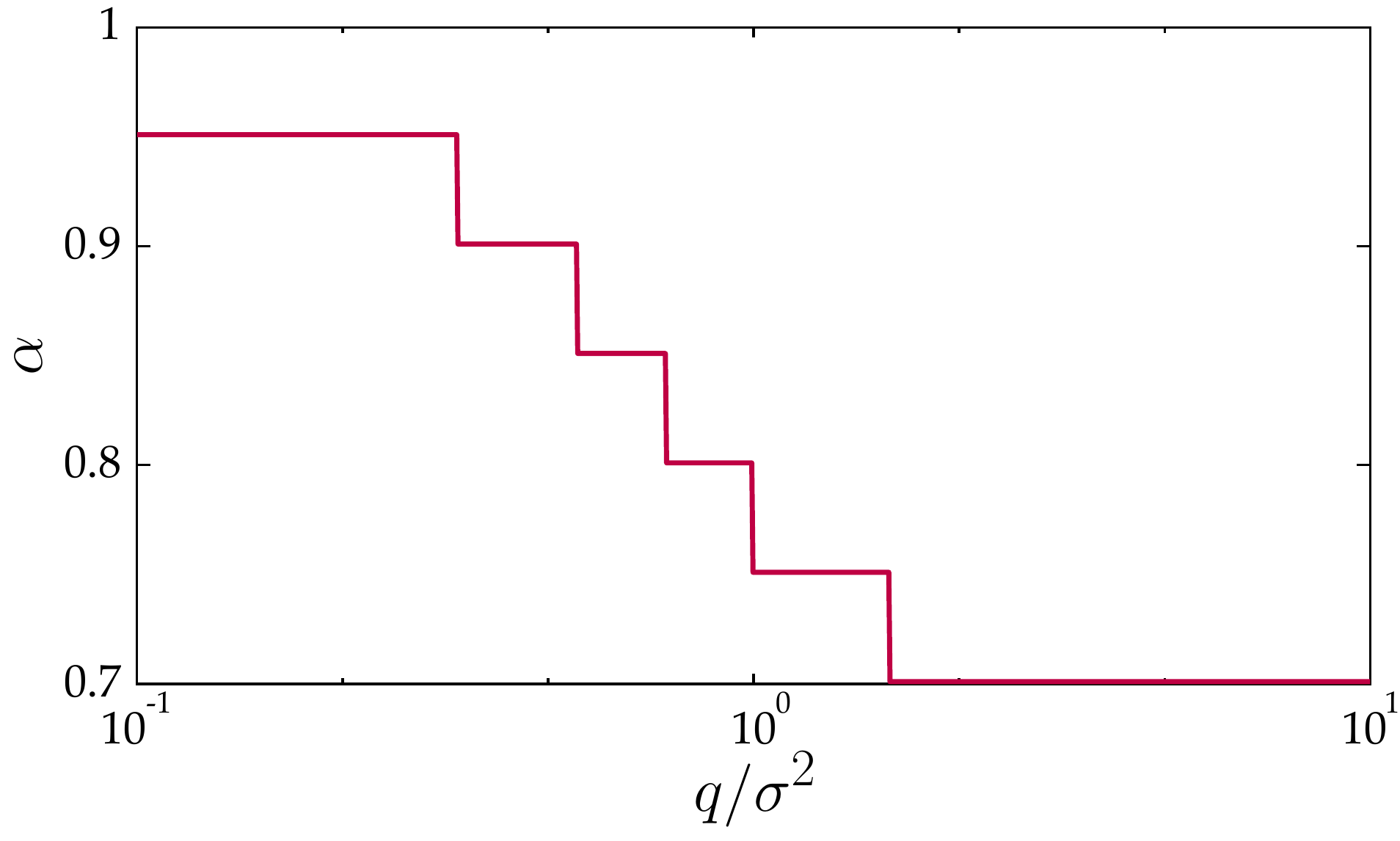}%
\vspace{-2mm}
\caption{\small $\alpha$ corresponding to $\kappa_n \approx 1$ against $q/\sigma^2$}
\label{fig:alpha_vs_q}
\vspace{-5mm}
\end{figure}

Fig. \ref{fig:kappa_vs_alpha}\textit{B} shows the plot of $\kappa_n$ vs. $\alpha$, which exhibits a similar increasing trend, but eventually exceeds $1$. This result implies that with a careful choice $\alpha$, it is possible to achieve $\kappa_n < 1$, and hence obtain lower variance than that of the standard MT estimate. Fig. \ref{fig:alpha_vs_q} illustrates this statement by showing the value of $\alpha$ for which $\kappa_n \approx 1$ vs. $q_n/\sigma^2$. For models with high temporal dependence (i.e., $\alpha$ close to $1$), it is possible to achieve $\kappa_n < 1$ and hence reduce the estimator variance, due to the increase in the weight of data pooled from adjacent windows, even for small values of $q_n/\sigma^2$ (i.e., low SNR). However, this reduction in variance comes with the cost of increasing the bias. On the contrary when the data across windows have low temporal dependence (i.e., $\alpha \ll 1$), it is only possible to achieve a reduction in variance for high values of $q_n/\sigma^2$ (i.e., high SNR). This  is due to the fact that at low SNR with low temporal dependence, pooling data from adjacent windows is not beneficial in reducing the variance in a particular window, and indeed can result in higher bias.

\textit{\textbf{Remark 3.}} Even though the parameter $\alpha$ is estimated in a data-driven fashion, it can be viewed as a tuning parameter controlling the bias-variance trade-off, given the foregoing discussion. For a given SNR, fixing $\alpha$ at a small value can help reduce the variance but with a cost of increasing bias, and vice versa. In light of this observation, Fig. \ref{fig:alpha_vs_q} can be thought of as a guideline for choosing $\alpha$ to achieve $\kappa_n \approx 1$, so that the estimator is nearly unbiased, and achieves a lower variance than that of the standard overlapping MT estimator due to the noise suppression virtue of the state-space model. Although we focused on the case of flat spectrum in the foregoing discussion, it is possible to numerically compute these trade-off curves for more general cases, given the general expressions for $\kappa_n(f)$ and $\mu_n(f)$ given in Appendix \ref{A1}.

\textit{\textbf{Remark 4.}} Extending Theorems 1 and 2 to the {\sf log-DBMT} algorithm is not straightforward, due to the high nonlinearity of the underlying state-space models. However, under the common Gaussian approximation of the log-posterior density, the well-known variance reduction property of the fixed interval smoother \cite{rauch65,anderson1979optimal} carries over to the estimator of $\log S_n(f)$. That is, the variance of the {\sf log-DBMT} estimate of $\log S_n(f)$ obtained using the state-space model is lower than that of the standard MT estimate which only uses the data within window $n$. This fact agrees with our earlier observation in Section \ref{sec:simulation} regarding the tightening of the confidence intervals for {\sf log-DBMT} as compared to the overlapping MT estimates.
\vspace*{-1mm}
\section{Concluding Remarks}\label{sec:conclusion}
\vspace*{-1mm}
Spectral analysis of non-stationary time series data poses serious challenges for classical non-parametric techniques, in which temporal smoothness of the spectral representations are implicitly captured using sliding windows with overlap. This widely-practiced approach tends to ignore the inherent smoothness of the data and is not robust {against} measurement/sampling noise. In this paper, we address these issues and provide an alternative to the sliding window spectrogram analysis paradigm. We propose two {semi-parametric} spectrogram estimators, namely the {\sf DBMT} and {\sf log-DBMT} estimators, by integrating techniques from MT analysis and Bayesian estimation. To this end, we explicitly model the temporal dynamics of the spectrum using a state-space model over the spectral features obtained by multitapering. Therefore our algorithms inherit the optimality features of both Bayesian estimators and MT analysis. 

Our algorithms admit efficient and simple implementations, thanks to the Expectation-Maximization algorithm and the well-known fixed interval state-space smoothing procedure. Unlike existing approaches, our algorithms require no \emph{a priori} assumptions about the structure of the spectral representation and operate in a fully data-driven fashion. While both algorithms yield {spectral estimates} that are continuous in time, by design the {\sf DBMT} algorithm significantly suppresses the {measurement} noise in forming the spectrogram and the {\sf log-DBMT} algorithm mitigates {sampling} noise due to small observation length. We establish the performance gains provided by our algorithms through theoretical analysis of the bias-variance trade-off, as well as application to synthetic and real data from human EEG recording during sleep and ENF signals from audio recordings.        

\vspace{-4mm}
\appendices
\section{Proofs of Theorems 1 and 2}
\label{A1}
\vspace{-1mm}
\label{pfthm1}
{Recall from Eq. (\ref{eq:xn}) that the $k$th eigen-coefficient estimate at window $n$ can be written in terms of the observed data as:
\vspace{-1mm}
\begin{align}
\label{eq:xnN}
    \widehat{\mathbf{x}}_{n|N}^{(k)} = \sum_{s=1}^{N}\mathbf{ \Lambda}^{|s-n|} \boldsymbol{\Gamma} \mathbf{F}_s^{H}\widetilde{\mathbf{y}}_s^{(k)},
\end{align}

\vspace{-2mm}
\noindent where $\widetilde{\mathbf{y}}_s^{(k)} = \mathbf{u}^{(k)}\odot \widetilde{\mathbf{y}}_s = \mathbf{U}^{(k)}\widetilde{\mathbf{y}}_s$.} Given that $\mathbf{Q}$ is a diagonal matrix with elements $q(f_j)$, $j=1,2,\cdots,J$, it can be shown that {$\mathbf{\Sigma}_{\infty}$}, $\mathbf{\Lambda}$ and $\mathbf{\Gamma}$ are also diagonal matrices. Denoting the elements of {$\mathbf{\Sigma}_{\infty}$}, $\mathbf{\Gamma}$ and $\mathbf{\Lambda}$, respectively by $\tau(f)$, $\eta(f)$ and $\gamma(f)$, for $f \in \{f_1,f_2,\cdots, f_J\}$, we have $
\eta(f) = \frac{\alpha^2\tau(f)+q(f)}{1+rW(\alpha^2\tau(f)+q(f))}$.

\vspace{-1mm}
\subsection{Proof of Theorem 1}
First note that Eq. (\ref{eq:z-ss}) implies that $\mathbbm{E}[{dz_n(f)dz_{n+t}^{*}(f')}] = \alpha^t D(f)\delta(f-f')dfdf'$. By invoking the Cram\'{e}r representation, the covariance of the data tapered by the $k$th and $l$th dpss sequences can be expressed as: 
\vspace{-1mm}
\begin{align}
&\resizebox{\columnwidth}{!}{$\displaystyle \mathbbm{E}\left[(\widetilde{\mathbf{y}}_{s}^{(k)})_j (\widetilde{\mathbf{y}}_{s'}^{(l)})_{j'}^*\right] = \alpha^{|s-s'|}\int_{-1/2}^{1/2}U_k(f_j-\beta)D(\beta)U_l^{*}(f_{j'}-\beta)d\beta$} \nonumber \\[-5pt]
& \qquad \quad \ \ \ \ \resizebox{0.67\columnwidth}{!}{$\displaystyle + \int_{-1/2}^{1/2}U_k(f_j-\beta)\sigma^2\delta(s-s')U_l^{*}(f_{j'}-\beta)d\beta$}.
\end{align}
\vspace{-1.5mm}
\noindent From Eqs. (\ref{est:dbmt}) and (\ref{eq:xnN}), we get:
\vspace{-1mm}
\begin{align}
\resizebox{.89\columnwidth}{!}{$\displaystyle \widehat{D}_{n|N}(f_j) = \frac{\eta^2(f_j)}{K}\sum_{k=1}^{K}\sum_{s=1}^{N}\sum_{s'=1}^{N} \gamma(f_j)^{|s-n|}\gamma(f_j)^{|s'-n|}(\widetilde{\mathbf{y}}_{s}^{(k)})_j (\widetilde{\mathbf{y}}_{s'}^{(k)})_{j}^*.$}
\end{align}
Taking the expectation of both sides and after some simplification, one arrives at:
\begin{align}
\label{ub}
& \resizebox{\columnwidth}{!}{$\displaystyle \mathbbm{E}[\widehat{D}_{n|N}(f_j)]=\frac{\eta^2(f_j)}{K}\sum_{k=1}^{K}\sum_{s=1}^{N}\sum_{s'=1}^{N}  \gamma(f_j)^{|s-n|} \gamma(f_j)^{|s'-n|} \mathbbm{E} \left[(\widetilde{\mathbf{y}}_{s}^{(k)})_j (\widetilde{\mathbf{y}}_{s'}^{(k)})_j^*\right]$} \nonumber \\[-4pt]
& \qquad \quad \quad \ =\bigg[\eta(f_j)^2\sum_{s=1}^{N}\sum_{s'=1}^{N} \gamma(f_j)^{|s-n|} \gamma(f_j)^{|s'-n|} \alpha^{|s-s'|}\bigg]  \times \nonumber \\[-5pt]
& \qquad \qquad \qquad \frac{1}{K}\sum_{k=1}^{K}\int_{-1/2}^{1/2} U_k(f_j-\beta) D(\beta) U_k^{*}(f_j-\beta)d\beta \nonumber \\[-5pt]
& \qquad \qquad \quad \ + \bigg[\eta(f_j)^2\sum_{s=1}^{N} \gamma(f_j)^{2|s-n|} \bigg] \sigma^2.
\end{align}
Using the orthogonality of the PSWFs as in \cite{lii2008}, and using the fact that $D(\beta) \leqslant \sup_f D(f), \forall \beta$, we get:
\begin{align}
\big|\mathbbm{E}[\widehat{D}_{n|N}&(f)]- \kappa_n(f) D(f) \big| \leqslant \kappa_n(f)(\sup_{f} \{D(f)\}-D(f)) \times  \nonumber \\[-5pt]
&\bigg(1-\frac{1}{K}\sum_{k=1}^{K}\lambda_k\bigg) + \mu_n(f) \sigma^2 +\kappa_n(f) o(1),
\end{align} 
\vspace{-4mm}
\[
\text{where } \kappa_n(f) := \eta(f)^2\sum_{s=1}^{N}\sum_{s'=1}^{N} \gamma(f)^{|s-n|} \gamma(f)^{|s'-n|} \alpha^{|s-s'|},
\]
$
\text{and }\mu_n(f) := \eta(f)^2\sum_{s=1}^{N} \gamma(f)^{2|s-n|},
$
for $f \in \{f_1,f_2,\cdots, f_J\}$. Using the triangle inequality, the bound of Theorem \ref{theorem1} on $\big|\mathbbm{E}[\widehat{D}_{n|N}(f)]- D(f) \big|$ follows. \QEDB

\vspace{-3mm}
\subsection{Proof of Theorem 2}
Using the notation of Appendix \ref{pfthm1}, we have:
\begin{align}\label{eq:cov}
&{\sf Cov}\left\{\widehat{D}_{n}^{(k)}(f),\widehat{D}_{m}^{(l)}(f')\right\} = \nonumber \\[-5pt]
&\eta(f)^4\!\!\!\!\sum_{s,s',t,t'=1}^{N}\!\!\!\gamma(f)^{|s-n|}\gamma(f)^{|s'-n|} \gamma(f')^{|t-m|} \gamma(f')^{|t'-m|}\times \nonumber \\[-5pt]
&\resizebox{.82\columnwidth}{!}{$\displaystyle \bigg[ \alpha^{|s-t|} \alpha^{|s'-t'|} \int\int U_k(f-\beta) U_l(f'+\beta) U_k(-\beta'-f)\times$} \nonumber 
\\[-3pt] 
&\!\resizebox{.89\columnwidth}{!}{$\displaystyle U_l(\beta'-f') (D(\beta)+\sigma^2\delta(s-t)) (D(\beta')+\sigma^2\delta(s'-t')) d\beta d\beta'$} + \nonumber\\[-3pt]
& \resizebox{.82\columnwidth}{!}{$\displaystyle\alpha^{|s-t'|} \alpha^{|s'-t|}\int\int U_k(f-\beta) U_l(\beta-f') U_k(-\beta'-f)\times$}  \nonumber \\[-5pt]
&\!\!\resizebox{.9\columnwidth}{!}{$\displaystyle U_l(\beta'+f')(D(\beta)+\sigma^2\delta(s-t')) (D(\beta')+\sigma^2\delta(s'-t)) d\beta d\beta'\bigg]$}
\end{align}
Note that we have omitted the integral limits, as they are understood to be same as in (\ref{eq:cramer}) henceforth. After summing over all tapers and rearranging the summations, the first expression within the brackets in {(\ref{eq:cov})} becomes:
\begin{align}\label{eq:cov2}
 &\resizebox{.98\columnwidth}{!}{$\displaystyle \int\!\!\int\bigg[\!\eta(f)^2\!\sum_{s,t=1}^{N}\! \gamma(f)^{|s-n|}\! \gamma(f')^{|t-m|} \!\alpha^{|s-t|}\! \bigg]\!(D(\beta)\!+\!\sigma^2\delta(s\!-\!t)) \times$} \nonumber \\[-6pt]
&\resizebox{.98\columnwidth}{!}{$\displaystyle \bigg[\!\eta(f)^2\!\sum_{s',t'=1}^{N}\!\gamma(f)^{|s'-n|}\!\gamma(f')^{|t'-m|}\!\alpha^{|s'-t'|}\!\bigg]\!(D(\beta')\!+\!\sigma^2\delta(s'\!-\!t')) \times$} \nonumber \\[-6pt] 
&\resizebox{0.89\columnwidth}{!}{$\displaystyle \sum_{k=1}^{K}U_k(f\!-\!\beta) U_k(\!-\!\beta'\!-\!f)\!\sum_{l =1}^{K}U_l(f'\!+\!\beta) U_l(\beta'\!-\!f') d\beta d\beta'.$}
\end{align}
Let \vspace*{-4mm}
\begin{align}
 \mathbbm{A}(n,m,f) &:= \bigg[\eta^2(f) \sum_{s,t=1}^{N}\gamma(f)^{|s-n|} \gamma(f')^{|t-m|} \alpha^{|s-t|} \bigg]D(f) \nonumber \\[-5pt]
 & \qquad +\bigg[ \eta(f)^2 \sum_{s=1}^{N} \gamma(f)^{2|s-n|} \bigg]\sigma^2
\end{align}

Using the Schwarz inequality, the integral in Eq. (\ref{eq:cov2}) can be bounded by:
\begin{align}
\label{ubvar}
 &\!\!\resizebox{.9\columnwidth}{!}{$\displaystyle\Bigg[\int\!\!\int \bigg|\sum_{k=1}^{K}U_k(f-\beta) U_k(-\beta'-f)\bigg|^2  
  \mathbbm{A}(n,m,\beta) \mathbbm{A}(n,m,\beta') d\beta d\beta'$} \nonumber\\ 
  &\!\!\resizebox{0.89\columnwidth}{!}{$\displaystyle\times \int\!\!\int \Bigg|\sum_{l =1}^{K}U_l(f'+\beta) U_l(\beta'-f')\Bigg|^2 
   \mathbbm{A}(n,m,\beta) \mathbbm{A}(n,m,\beta') d\beta d\beta'\Bigg]^{1/2}$},
\end{align} 
Using bounds on the convolutions of PSWFs {from} \cite{lii2008}, and upper bounding $\mathbbm{A}(n,m,\beta)$ by $\sup_f \{ \kappa_n(f) D(f) + \mu_n(f) \sigma^2\}$, the statement of the theorem on the variance of the {\sf DBMT} estimate $\widehat{D}_n(f)$ follows. \QEDB

\section{Characterization of $\kappa_n(f)$: bounds and parameter dependence}
\label{A3}
\subsection{Lower and Upper Bounds on $\kappa_n(f)$}
Consider the scenario where $q(f) = q$, i.e., flat spectrum. Then, the dependent of $\gamma(f)$ and $\kappa_n(f)$ on $f$ is suppressed. We have the following bound on $\kappa_n$:
\begin{prop}
\label{prop:kappa}
For $0< \gamma, \alpha <1 $, the quantity $\kappa_n$ can be bounded as:
\begin{align}
\nonumber  \resizebox{\columnwidth}{!}{$\displaystyle \Bigg|\kappa_n - \bigg(1-\frac{\gamma}{\alpha}\bigg)^2 \bigg[ \frac{1+\alpha \gamma - 2(\alpha \gamma)^N}{1-\alpha \gamma}T_0 +\frac{\gamma}{(1-\gamma)^2}\bigg]\Bigg| \leqslant \bigg(1-\frac{\gamma}{\alpha}\bigg)^2 \frac{\gamma}{(1-\gamma)^2},$}
\end{align} 
where $\displaystyle T_0 := \frac{1+\gamma^2 -\gamma^{2n}-\gamma^{2(N-n+1)}}{1-\gamma^2}$.
\end{prop}

\begin{IEEEproof}
To get an upper bound on $\kappa_n$, we rewrite the expression defining $\kappa_n$ as:  
\begin{align}
\nonumber \resizebox{\columnwidth}{!}{$\displaystyle \kappa_n =\eta^2\sum_{s=1}^{N}\sum_{s'=1}^{N} \gamma^{|s-n|} \gamma^{|s'-n|} \alpha^{|s-s'|} =\eta^2\!\!\!\!\!\!\sum_{t=-N+1}^{N-1} \alpha^{|t|} \sum_{s=1}^{N} \gamma^{|s-n|} \gamma^{|s-t-n|}.$}
\end{align}
Now, let us define $T_t := \sum_{s=1}^{N} \gamma^{|s-n|} \gamma^{|s-t-n|}$. Then it can be verified that:
\begin{small}
\begin{align}
T_{t+1}
\begin{cases}
=\gamma T_t, \text{  when } t\geqslant N-n \\
\leqslant \gamma T_{t} + \gamma^{t+1}, \text{  when } 0 \leqslant t< N-n \\  
\end{cases}
\end{align}
\end{small}
and
\begin{small}
\begin{align}
T_{t-1}  
\begin{cases}
=\gamma T_t, \text{  when }  t \leqslant -n+1 \\
\leqslant \gamma T_{t} + \gamma^{|t-1|}, \text{  when } -n+1 < t < 0.  \\  
\end{cases}
\end{align}
\end{small}
Also, we have:
\begin{small}
\begin{align}
\label{sum1}
\sum_{t=0}^{N-1} \alpha^{t}T_t  &\leqslant \sum_{t=0}^{N-1} \alpha^{t}\gamma^t T_0 + \sum_{t=1}^{N-n-1} t \gamma^t + \sum_{t=N-n}^{N-1} (N-n)\gamma^t    \nonumber \\[-3pt]
&\leqslant \frac{1-(\alpha \gamma)^N}{1-\alpha \gamma}T_0 + \frac{\gamma}{(1-\gamma)^2}.
\end{align}
\end{small}
Similarly, we have:
\begin{small}
\begin{align}
\label{sum2}
\sum_{t=-N+1}^{0} \alpha^{|t|}T_t \leqslant \frac{1-(\alpha \gamma)^N}{1-\alpha \gamma}T_0 + \frac{\gamma}{(1-\gamma)^2},
\end{align}
\end{small}
which along with (\ref{sum1}) leads to the claimed upper bound.
For the lower bound, we use the fact that
\begin{small}
\begin{align}
\gamma T_t = 
\begin{cases}
\leqslant T_{t+1} \text{  when  }  t\geqslant0\\
\leqslant T_{t-1} \text{  when  }  t\leqslant0\\
\end{cases},
\end{align}
\end{small}
which implies $\sum_{t=0}^{N-1} \alpha^{t}T_t \geqslant  \frac{1-(\alpha \gamma)^N}{1-\alpha \gamma}T_0$ and $\sum_{t=-N+1}^{0} \alpha^{|t|}T_t \geqslant \frac{1-(\alpha \gamma)^N}{1-\alpha \gamma}T_0$. Using the latter lower bounds for $\kappa_n$ yield the claimed lower bound.
\end{IEEEproof}

\vspace{-3mm}
\subsection{Relation between $\kappa_n$ and $\mathbf{Q}$}
\label{A4}

The expressions for $\kappa_n(f)$ and $\mu_n(f)$ in Appendix \ref{A1} depend on $\gamma(f)$ and $\alpha$. But $\gamma(f)$ itself depends on $q(f)$ and $\alpha$, and it is not straightforward to give a closed-form expression of $\gamma(f)$ merely in terms of $q(f)$ and $\alpha$, since it requires computation of the filtered error covariance matrix $\mathbf{\Sigma}_{n|n}$ given $\mathbf{Q}$.  Again, by invoking the stead-state approximation, and defining $\mathbf{\Sigma}_{n|n-1} =: \mathbf{\Sigma}, \forall n = 1,2, \cdots,N$, the matrix $\boldsymbol{\Sigma}$ can be obtained by solving the following algebraic Riccati equation:
\begin{align}
\label{Ric}
    \mathbf{\Sigma} = \alpha^2\mathbf{\Sigma} - \alpha^2 \mathbf{\Sigma} \mathbf{F}_n^{H}\big( \sigma^2\mathbf{I} + \mathbf{F}_n\mathbf{\Sigma}\mathbf{F}_n^{H} \big)^{-1} \mathbf{F}_n \mathbf{\Sigma} + \mathbf{Q}
\end{align}
and thereby the steady-state error covariance matrix $\boldsymbol{\Sigma}_{n|n} =: \boldsymbol{\Sigma}_\infty$ is given by $\mathbf{\Sigma}_\infty = \frac{1}{\alpha^2}(\mathbf{\Sigma} - \mathbf{Q})$.

Although this procedure {can} be carried out numerically, in general it is not possible to solve the Riccati equation to get a closed-form expression for arbitrary $\mathbf{Q}$. In order to illustrate the explicit dependent of $\kappa_n(f)$ on the state-space parameters, we consider the case of flat spectrum where $\mathbf{Q} = q\mathbf{I}$. In this case, it can be shown that $\mathbf{\Sigma} = \zeta \mathbf{I}$ for some $\zeta > 0$ and the matrix equation (\ref{Ric}) reduces to a simpler scalar equation for $\zeta$, given by:
\begin{align}
\resizebox{.80\columnwidth}{!}{$\displaystyle \frac{\zeta}{\sigma^2} = \alpha^2 \frac{\zeta}{\sigma^2}\left[ 1 -  \frac{\zeta}{\sigma^2} \bigg(1 - \frac{rW (\zeta/\sigma^2)}{rW(\zeta/\sigma^2) +1}\bigg)rW\right] + \frac{q}{\sigma^2}$}.
\end{align}
Following simplification, a quadratic equation for $\zeta$ results, and since $\zeta \geqslant 0$, the positive solution for $\zeta$ is given by:
\begin{align}
\!\!\!\!\resizebox{.92\columnwidth}{!}{$\displaystyle \frac{\zeta}{\sigma^2} =\frac{1}{2rW} \left[ -\left(1 - \alpha^2 - rW\frac{q}{\sigma^2}\right) +\sqrt{\left(1-\alpha^2- rW\frac{q}{\sigma^2}\right)^2 + 4rW\frac{q}{\sigma^2}}\right].$}
\end{align}
Then, $\gamma$ can be computed as $\gamma = \frac{1}{\alpha} \big( 1-\frac{q/\sigma^2}{\zeta/\sigma^2}\big)$. Using these expressions for $\gamma$ and $\alpha$, the functions $\kappa_n$ and $\mu_n$ can be computed. Fig. \ref{fig:kappa_vs_alpha}\textit{B} shows the upper and lower bounds on $\kappa_n$ evaluated for $n = 50$ and different values of $\alpha$ for $q/\sigma^2 = 10$. The upper and lower bounds are simple functions of $\alpha$ and $q/\sigma^2$ and can be used to further inspect the performance trade-offs of the {\sf DBMT} algorithm with respect to the state-space model parameters.

\section*{Acknowledgment}
We would like to thank Adi Hajj-Ahmad and Min Wu for providing us the ENF data from \cite{Wu2013}.

\ifCLASSOPTIONcaptionsoff
  \newpage
\fi


\bibliographystyle{IEEEtran}
\bibliography{DBMT_Final}

%








\end{document}